\documentclass[manuscript,nonacm]{acmart}
\usepackage{multirow}
\usepackage{multicol}
\usepackage{arydshln}
\usepackage{longtable}
\usepackage{array}
\usepackage{colortbl}
\usepackage{xcolor}
\usepackage{soul}
\usepackage{changepage}
\usepackage{enumitem}
\usepackage{arydshln}
\usepackage{tikz}

\definecolor{lightblue}{HTML}{b0dcf4} 
\definecolor{lightgreen}{HTML}{d8ecc4} 
\definecolor{lightyellow}{HTML}{f8f4b0} 
\definecolor{lightpurple}{HTML}{decee5} 
\definecolor{DR_yellow}{HTML}{ffde17} 
\definecolor{DR_green}{HTML}{8bc640} 
\definecolor{DR_purple}{HTML}{9864ac} 
\definecolor{DR_bluegreen}{HTML}{c9ebe7} 
\definecolor{DR_yellowgreen}{HTML}{f6f6cf} 
\definecolor{panels}{HTML}{9c206e}
\newcommand{\highlight}[2]{\colorbox{#1}{#2}}

\definecolor{mediumgray}{HTML}{808080}

\definecolor{VA}{HTML}{FFDFB5} 
\definecolor{design}{HTML}{F7F8A7}
\sethlcolor{design} 
\newcolumntype{M}{>{\ttfamily\small}p{14cm}}
\newcolumntype{L}{>{\raggedright\arraybackslash}p{2.7cm}}

\definecolor{custompurple}{HTML}{a275ab}
\definecolor{custompink}{HTML}{f27d9b}

\newcolumntype{S}{>{\ttfamily\small}p{8.5cm}}
\newcolumntype{X}{>{\columncolor{VA}\ttfamily\small}p{8.5cm}}
\newcommand{\pquotes}[1]{\textcolor[gray]{0.35}{\textit{#1}}}
\newcommand{\tquotes}[1]{{\textit{#1}}}

\def\eg{\emph{e.g., }}

\newcommand{\am}[1]{\textcolor{black}{#1}}

\newcommand{\lowthr}{2.50}
\newcommand{\highthr}{3.50}
\newcommand{\heat}[1]{%
  \begingroup\def\val{#1}%
  \ifdim \val pt < \lowthr pt
    \cellcolor{red!15}\val%
  \else\ifdim \val pt > \highthr pt
    \cellcolor{green!20}\val%
  \else
    \cellcolor{orange!30}\val%
  \fi\fi\endgroup
}

\newcommand{\susLow}{60}
\newcommand{\susHigh}{70}

\newcommand{\heatSUS}[1]{%
  \begingroup\def\val{#1}%
  \ifdim \val pt < \susLow pt
    \cellcolor{red!15}\val%
  \else\ifdim \val pt < \susHigh pt
    \cellcolor{orange!30}\val%
  \else
    \cellcolor{green!20}\val%
  \fi\fi\endgroup
}

\AtBeginDocument{%
  }

\setcopyright{acmlicensed}
\copyrightyear{2018}
\acmYear{2018}
\acmDOI{XXXXXXX.XXXXXXX}
\acmConference[DIS '26]{ACM Designing Interactive Systems Conference}{  13 – 17 June 2026}{Singapore}

\begin{document}

\title[Conversational Agents in Behavioral Sleep Medicine: Designing Self-Report and Analytics Tools]{Conversational Agents in Behavioral Sleep Medicine: \\ Designing Self-Report and Analytics Tools}

\author{Amama Mahmood}
\email{amama.mahmood@jhu.edu}

\affiliation{%
  \institution{The Johns Hopkins University}
  \streetaddress{3400 N. Charles St}
  \city{Baltimore}
  \state{Maryland}
  \country{USA}
  \postcode{21218}
}

\author{Nadia Kim}
\email{bkim94@jhu.edu}
\affiliation{%
  \institution{The Johns Hopkins University}
  \streetaddress{3400 N. Charles St}
  \city{Baltimore}
  \state{Maryland}
  \country{USA}
  \postcode{21218}
}

\author{Honghao Zhao}
\email{hzhao78@jhu.edu}
\affiliation{%
  \institution{The Johns Hopkins University}
  \streetaddress{3400 N. Charles St}
  \city{Baltimore}
  \state{Maryland}
  \country{USA}
  \postcode{21218}
}

\author{Molly E. Atwood}
\email{matwood4@jhmi.edu}
\affiliation{%
  \institution{Department of Psychiatry and Behavioral Sciences, Johns Hopkins University School of Medicine}
  \city{Baltimore}
  \state{Maryland}
  \country{USA}
}

\author{Luis F. Buenaver}
\email{lbuenav1@jhmi.edu}
\affiliation{%
  \institution{Department of Psychiatry and Behavioral Sciences, Johns Hopkins University School of Medicine}
  \city{Baltimore}
  \state{Maryland}
  \country{USA}
}

\author{Michael T. Smith}
\email{msmith62@jhmi.edu}
\affiliation{%
  \institution{Department of Psychiatry and Behavioral Sciences, Johns Hopkins University School of Medicine}
  \city{Baltimore}
  \state{Maryland}
  \country{USA}
}

\author{Chien-Ming Huang}
\email{chienming.huang@jhu.edu}
\affiliation{%
  \institution{The Johns Hopkins University}
  \streetaddress{3400 N. Charles St}
  \city{Baltimore}
  \state{Maryland}
  \country{USA}
  \postcode{21218}
}

\renewcommand{\shortauthors}{Mahmood et al.}

\begin{abstract}

The sleep diary is a widely used clinical tool for understanding and treating sleep disorders in \am{Behavioral Sleep Medicine (BSM)}; however, low patient compliance and limited capture of contextual information constrain its effectiveness and leave specialists with an incomplete picture of patients’ sleep-related behaviors. 
\am{In this work, we explore  conversational agents (CAs) as an alternative to traditional diary methods by designing a voice-based sleep diary and a specialist-facing analytics tool, and using them as design probes to understand how CAs might support BSM more broadly.} 
Our multi-stage study with specialists comprised: (1) interviews to identify shortcomings of current text-based diaries, (2) iterative co-design of the conversational diary and \am{analytics tool}, and (3) focus groups to examine broader opportunities for CAs in BSM.
\am{This work offers empirical insights into how specialists envision CAs in clinical care and outlines design implications for integrating them into existing self-report practices and behavioral interventions.}


\end{abstract}



\renewcommand{\thefootnote}{}
\footnotetext{This is the author’s version of the work. It is posted here for your personal use. Not for redistribution.}

\renewcommand{\thefootnote}{\arabic{footnote}}


\begin{teaserfigure}
  \includegraphics[width=\textwidth]{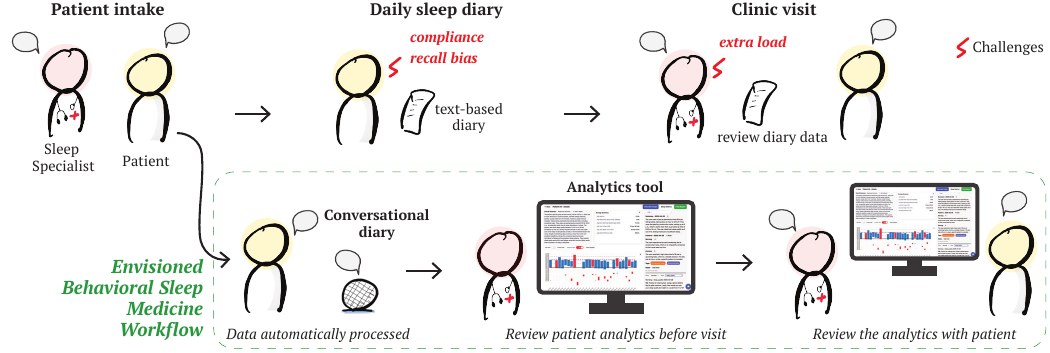}
  \caption{ \am{In this work, we explore how conversational agents (CAs) can serve as an alternative to traditional text-based sleep diary methods to support Behavioral Sleep Medicine (BSM). We developed a conversational sleep diary and a specialist-facing analytics tool as design probes to understand the role of CAs in BSM through the lens of sleep specialists.}}
  \Description{This figure shows the traditional behavioral sleep medicine workflow and the envisioned behavioral sleep medicine workflow.  
  In the first diagram, there are two stick figures: the one on the left represents the sleep specialist, and the one on the right represents the patient. They are having a conversation, with the title above labeled patient intake. The next step of the workflow branches into two arrows. The top arrow points to the traditional workflow, which shows a stick figure representing the patient. To the right, there is a stick-drawn sheet of paper labeled text-based diary. Next to it, in red text, the challenges are written as compliance and recall bias. The title above this step is Daily sleep diary.  
  The arrow on the right side points to two stick figures: the one on the left is the sleep specialist and the one on the right is the patient. Between them is a stick-drawn sheet of paper with the label review diary data. Between the two figures, red text indicates the challenge as extra load. The title above this drawing is clinic visit.  
  The bottom arrow points to a sequence of three drawings from left to right, all under the shared title Envisioned Behavioral Sleep Medicine Workflow.  
  The first drawing shows a stick figure patient and a stick figure conversational agent, engaged in dialogue. Below it is the caption data automatically processed, and above it the title conversational diary.  
  The first drawing points to the second drawing, which depicts a stick figure sleep specialist on the left and a stick-drawn computer on the right, displaying the designed interface. Below it is the caption review patient analytics before visit, and above it the title analytics tool.  
  The second drawing points to the third drawing, which shows two stick figures: the one on the left is the sleep specialist, and the one on the right is the patient. They are conversing, with a stick-drawn computer displaying the designed interface placed between them. Below it is Review the analytics with the patient.  
  }
  \label{fig:teaser}
\end{teaserfigure}



\maketitle

\section{Introduction}

About one-third of people struggle to fall asleep or stay asleep \cite{resmed2025Global}. Sleep disorders such as insomnia, sleep apnea, and narcolepsy are prevalent globally; for instance, 16.2\% of people are estimated to have insomnia \cite{benjafield2025estimation}. Inadequate sleep is linked to chronic conditions such as cardiovascular disease, diabetes, and obesity, as well as impaired mental health, reduced productivity, and higher accident risk \cite{altevogt2006sleep, shah2025effects, lim2023need}. These consequences highlight sleep health as both a clinical and public health priority.
\am{This has led to a growing emphasis on Behavioral Sleep Medicine (BSM), which addresses sleep disorders through evidence-based behavioral interventions \cite{ong2018concept}.  
Among behavioral interventions, Cognitive Behavioral Therapy for Insomnia (CBT-I) has shown long-term effectiveness, either alone or alongside other treatments \cite{rossman2019cognitive, van2019cognitive, anwar2025effect}, and is considered the first-line treatment for chronic insomnia by major clinical guidelines \cite{edinger2021behavioral, qaseem2016management}. CBT-I is a multicomponent intervention that combines behavioral strategies (sleep restriction, stimulus control) with cognitive techniques to address maladaptive sleep patterns \cite{morin2009cognitive, perlis2005cognitive}.
To diagnose and understand sleep patterns, specialists employ various tools including actigraphy (objective monitoring of sleep-wake patterns) and sleep diaries (subjective self-reports) \cite{ancoli2003role, buysse2006recommendations, kushida2001comparison}.
}

\am{Indeed, the sleep diary is an integral and widely used component of CBT-I \cite{carney2012consensus, sateia2014international, buysse2006recommendations}---used to establish baseline sleep patterns, guide behavioral interventions, and monitor adherence over time \cite{morin2009cognitive,perlis2005cognitive,edinger2021behavioral,bootzin1972behavioral}.
A recent systematic review of 241 CBT-I trials found that sleep diaries were included in 192 of 210 treatment arms, representing 91 percent \cite{furukawa2024components}. 
Sleep diaries enable behavioral sleep specialists---referred to hereafter as sleep specialists or simply specialists\footnote{Behavioral sleep specialists are clinicians with PhD or PsyD training in psychology or behavioral sciences, rather than medical doctors (MDs).}---to monitor relevant sleep metrics alongside objective tools such as actigraphy and use them to create interventions and treatment plans. The Consensus Sleep Diary (CSD), developed to standardize prospective sleep self-reporting, is now widely adopted in both clinical practice and research \cite{carney2012consensus}.}

Clinically, sleep diaries are most often completed on paper, with occasional use of digital formats such as RedCAP \cite{harris2009redcap}.
\am{Prior work points to various challenges faced by these sleep diaries such as low patient compliance as burden of completion falls on patients \cite{clegg2023real, thurman2018individual} and recall bias as patients frequently misremember or approximate key details of past experiences \cite{ibanez2018survey}. Moreover, while standardized formats of sleep diaries (\eg CSD \cite{carney2012consensus}) and objective tools (\eg actigraphy) are used to capture core sleep parameters reliably, they can omit contextual factors (\eg lifestyle, environment) and subjective states (\eg mood, fatigue, stress). However, these factors have shown to strongly  influence sleep health \cite{johnson2018environmental, billings2020physical, chang2020impact, irish2015role} and thus can create a gap in specialists' understanding of sleep behaviors.} Digital diaries were introduced to improve convenience \cite{mohr2017behavioral, almahmud2022sleepapps, arigo2025behavioral}, but most simply replicate the static CSD structure \cite{kristbergsdottir2023evaluating, arnardottir2022sleep}. As a result, they also remain limited in capturing context.
Specialists, meanwhile, face their own challenges. Diary data is often not available until sessions \cite{islind2019shift}, and analysis requires manual labor that extends into unpaid ``pajama time'' \cite{schmitz2022towards}. These inefficiencies persist in part because pen-and-paper diaries lack analytics tools. Although digital platforms offer analytics, they are primarily patient-facing---intended to boost engagement and self-monitoring \cite{almahmud2022sleepapps, philip2020covidagent, kuhn2016cbticoach, espie2012sleepio}---rather than designed to support specialists.

\am{To further understand the shortcomings of traditional sleep diaries and contextualize them within BSM practice through the lens of sleep specialists, we conducted formative interviews with sleep specialists.  Our empirical findings from interviews reinforced the shortcomings of traditional sleep diaries expressed in prior work, including low patient compliance due to burden, limited context capture due to recall bias, and increased burden on specialists due to manual data analysis (Fig. \ref{fig:interview-findings-workflow-shortcomings}).} 

Conversational agents (CAs)---such as voice assistants (VAs)\footnote{We use \textit{conversational agent (CA)} to refer to AI systems that interact via natural language, and \textit{voice assistant (VA)} when voice is the primary modality. We use the term CA as our system has both text and voice modality.}---\am{offer a promising alternative to traditional, static diaries in capturing essential sleep metrics as well as addressing these shortcomings};
\am{voice-based self-reporting in health and well being contexts can provide a more engaging and potentially less burdensome hands-free experience than conventional input methods \cite{maharjan2022difference, chen2024dozzz, sinha2022adherence, groninger2025voicecancer}. While static diary formats constrain users to fixed fields and predetermined questions, conversational interfaces allow users to express themselves more naturally and elaborately during self-report \cite{maharjan2022difference}.}
\am{Advances in large language models (LLMs) further extend this potential by enabling CAs to capture self-reports more accurately and consistently \cite{wei2024leveraging, xu2025exploring, mahmood2025voice, yang2024talk2care, li2024diaryhelper}, making them a suitable alternative to capture context on top of the essential metrics reliably}.
This motivates our first research question: \textit{\am{How can we design a voice-based conversational sleep diary to enable adaptive capture of patient sleep behaviors and context?}}  However, such conversational input inevitably produces unstructured data, raising the challenge of how specialists can make use of it without increasing their burden. Thus, our second research question becomes: \textit{\am{How can we design analytics tools
for sleep specialists to process unstructured diary conversations in meaningful and clinically relevant ways?}}  
Finally, we want to understand: \textit{\am{How do specialists envision the broader roles CAs could play in BSM?}}

\am{To address these questions, we  engaged sleep specialists in an iterative co-design process to develop an LLM-powered VA sleep diary and specialist-facing analytics tool. We used these tools as design probes to explore specialists’ visions for how conversational agents could support BSM more broadly through focus groups.  Through this design process, specialists envisioned not only enhancements to diary intake and data analytics but also ways CAs could support interventions before, during, and after treatment. Our work makes three contributions: }

\begin{enumerate} [leftmargin=*, nolistsep]
    \item \am{Empirical insights from specialists on shortcomings  of traditional diaries in clinical context} (Section \ref{sec:interviews}).
    \item \am{Design and implementation of two systems---an LLM-powered conversational sleep diary and a specialist-facing analytics tool---developed through iterative co-design with specialists and  used as design probes (Section \ref{sec:interface-design}).} 
    \item Empirical insights from specialists on how conversational agents could support BSM beyond diary intake \am{to support} behavioral sleep interventions in new ways (Section \ref{sec:co-design-workshops}).
\end{enumerate}

\section{Related Work}

\am{Objective tools such as polysomnography and actigraphy are used to assess sleep by capturing physiological patterns and sleep behaviors \cite{ancoli2003role, kushida2001comparison}. In clinical practice, these objective measures are complemented by subjective self-report instruments that capture patients' perceived sleep experience \cite{buysse2006recommendations, carney2012consensus}. In this work,} we review prior work on patient self-reporting, with focus on sleep diaries in BSM. 



\subsection{Self-Reported Data Collection in Healthcare}
Self-reported data collection is widely used across healthcare domains to monitor medication adherence, track symptoms, assess mental health, and record lifestyle factors such as diet, exercise, and sleep patterns \cite{stone2002capturing,shiffman2008ema,arigo2019history}. Traditionally, these data have been collected through paper forms, structured questionnaires, and spreadsheets, with more recent transitions to electronic case report forms within platforms such as REDCap \cite{bolger2003diary,harris2009redcap}. These approaches provide standardized templates that make data entry straightforward, but face several limitations. 
Traditional self-report tools often struggle to capture contextual details that influence health outcomes, such as environmental conditions, stressors, and behavioral patterns, because they rely on static question sets and lack adaptive follow-up or context-aware questioning \cite{bolger2003diary,mohr2014bit,riley2011models}. Maintaining long-term engagement also remains a challenge, as users frequently experience reporting fatigue, leading to reduced adherence over time \cite{smyth2003ema,bolger2003diary,kristbergsdottir2023evaluating}. When multiple entries are completed retrospectively, recall bias---inability to remember past experiences accurately due to memory erosion---can further reduce data reliability \cite{shiffman2008ema,bolger2003diary}.

Recent advances in digital health have introduced mobile applications, wearable devices, and conversational agents to improve engagement and reduce participant burden \cite{mohr2017behavioral}. Digital tools have been used in chronic disease management, mental health monitoring, and lifestyle tracking, showing promise for supporting personalized care \cite{ghadi2025wearable,arigo2025behavioral}. However, effectively integrating contextual information and ensuring data quality remain unresolved challenges \cite{ghadi2025wearable,arigo2025behavioral,canali2022wearables}.

\subsection{Self-Reported Data Collection in Behavioral Sleep Medicine}

Within the broader healthcare landscape, sleep health represents a particularly critical domain where improved self-reporting methods are needed. Poor sleep is strongly associated with negative physical and mental health outcomes \cite{walker2021sleep,benjafield2019osa}, making accurate and context-rich data essential for effective intervention. In behavioral sleep medicine, self-reported tools---such as sleep diaries---are widely used to assess patient sleep patterns and guide treatment planning \cite{carney2012consensus}. Thus, in this work, we focus on patient self-reporting in behavioral sleep medicine. 

\subsubsection{Sleep diary---a self-report tool}
\am{Sleep diaries are widely used in BSM as monitoring and assessment tools and are an integral component of CBT-I---the first line treatment for chronic insomnia \cite{edinger2021behavioral,qaseem2016management,morin2009cognitive}. These diaries provide specialists with critical information such as bedtime, wake time, sleep onset latency, number and duration of awakenings, total sleep time, and perceived sleep quality \cite{carney2012consensus,ibanez2018survey}.}   Consensus Sleep Diary (CSD) \cite{carney2012consensus} is a widely adopted sleep diary. Traditionally administered using paper forms, spreadsheets, or structured electronic reporting systems, these diaries standardize data collection but struggle to capture contextual details such as lifestyle factors, stressors, and environmental conditions known to influence sleep \cite{bolger2003diary,johnson2018environmental,billings2020physical}. 
Moreover, maintaining long-term engagement is challenging; patient compliance often declines over time \cite{kristbergsdottir2023evaluating}, and retrospective “backfilling” introduces recall bias and reduces data reliability \cite{clegg2023real,shiffman2008ema}. Finally, manual data entry and analysis create additional burdens for both patients and specialists, limiting timely integration into clinical workflows \cite{blake2010online,vacaretu2019digitalsleep,sadhu2023dashboard}.

To address these limitations, recent research has explored a range of digital alternatives aimed at reducing participant burden and improving engagement \cite{mohr2017behavioral,almahmud2022sleepapps,arigo2025behavioral}. Mobile applications are the most widely adopted, often integrating visual summaries and basic analytics to provide feedback on key metrics such as sleep efficiency, average sleep duration, and time in bed \cite{almahmud2022sleepapps}. 
For example, during the COVID-19 pandemic, a smartphone-based virtual agent guided users through a one-week diary intake and visualized their sleep indicators \cite{philip2020covidagent}.
Although digital diaries offer advantages \cite{mohr2017behavioral, almahmud2022sleepapps, arigo2025behavioral}, most simply digitize the static consensus sleep diary (CSD) format \cite{kristbergsdottir2023evaluating, arnardottir2022sleep}. As such, they remain limited in capturing context and continue to face compliance and recall issues, since the reporting burden still rests on patients \cite{clegg2023real, kristbergsdottir2023evaluating}.
To reduce patient burden, wearables (\eg Fitbit) have been used in research to capture objective sleep measures. While diary and wearable averages align at the group level, discrepancies at the individual level---especially with frequent awakenings or poor perceived sleep---limit their clinical use, as validity remains unestablished \cite{jang2023chatbotfitbit,silva2007relationship}.

More recently, conversational systems have emerged as promising alternatives to static diary formats. In studies with children, voice-based diaries such as Dozzz showed higher engagement than text-based diaries in within-subject experiments \cite{chen2024dozzz}, while multimodal diaries combining voice and text inputs supported sustained reporting over a five-day field study \cite{chen2025multimodal}. Extending these insights to adults, a two-week feasibility study of an Alexa-based virtual assistant delivering CBT-I reported high engagement and positive feedback \cite{groninger2025voicecancer}. Similarly, other pilot studies suggest that voice-based conversational agents can enhance engagement and support self-reported sleep tracking \cite{almzayyen2022voiceagents}.
Collectively, these studies highlight the potential of conversational agents to address key limitations of traditional sleep diaries by enabling richer, more adaptive, and context-aware data collection. However, most evaluations remain short-term and outside clinical workflows, limiting their applicability to BSM \cite{mohr2017behavioral,mohr2014bit,riley2011models}. Prior work on patient-generated health data and specialist-facing tools also emphasizes that successful adoption requires alignment with specialist workflows and the active involvement of domain experts during design and evaluation \cite{islind2019shift,cerna2020changing,tiase2020integration,sadhu2023dashboard}. To address this gap, our study engages sleep specialists throughout the design and evaluation process to assess the feasibility of a conversational sleep diary and explore how conversational agents can extend beyond diary intake to support BSM more broadly.

\subsubsection{\am{Sleep diary data analytics tools}}

Despite their widespread use of diaries in BSM, analytics tools for sleep diary data remains limited. 
When available, analytics are often restricted to static charts summarizing basic metrics such as total sleep time, sleep efficiency, and wake after sleep onset \cite{blake2010online,almahmud2022sleepapps}. In many clinical settings, these summaries are still manually generated from paper records or electronic forms, making it difficult to track trends consistently across visits \cite{carney2012consensus,ibanez2018survey}.
To address these shortcomings, prior work has explored digital tools that pair diary intake with visual summaries for patient feedback \cite{philip2020covidagent} and mobile applications that integrate wearable data to enhance sleep tracking and engagement \cite{almahmud2022sleepapps}. Platforms such as \textit{CBT-I Coach}, \textit{SHUTi}, and \textit{Sleepio} further demonstrate that visual feedback can improve adherence and outcomes \cite{kuhn2016cbticoach,espie2012sleepio}. Yet most existing systems remain patient-facing and provide limited support for clinicians, who need integrated, longitudinal views of patient-reported data to guide treatment \cite{schmitz2022towards,sadhu2023dashboard,islind2019shift}. This gap not only increases clinician burden and the risk of error but also constrains personalized treatment planning \cite{vacaretu2019digitalsleep,schmitz2022towards,sadhu2023dashboard}.
\am{Our work addresses this need by engaging sleep specialists in the design of an analytics tool that aims to integrate diary metrics with contextual information grounded in existing clinical workflows to better support both individualized care and broader treatment planning.} 

\subsection{Potential of Large Language Models in Self-Reporting}

Recent studies demonstrate the potential of LLM-powered conversational agents for more natural and adaptive health data collection. For example, \textit{Talk2Care} improves data completeness through dynamic questioning \cite{yang2024talk2care}, \textit{MindfulDiary} shows that conversational journaling increases engagement and consistency in psychiatric care \cite{kim2024mindfuldiary}, and \textit{PhysioLLM} integrates wearable data with contextual cues to deliver personalized insights \cite{fang2024physiollm}.
Additionally, LLMs have shown potential to extract clinically relevant insights from free-text inputs and generate summaries to better integrate patient-reported data into healthcare workflows \cite{yang2024talk2care}. Collectively, these approaches suggest that LLMs can reduce reporting burden, capture richer context, and present data in more actionable ways. However, LLM-based methods have not yet been evaluated in clinical settings for behavioral medicine. Our study addresses this gap by assessing their feasibility and potential to support specialists in clinical settings.

\begin{figure}
  \includegraphics[width=\textwidth]{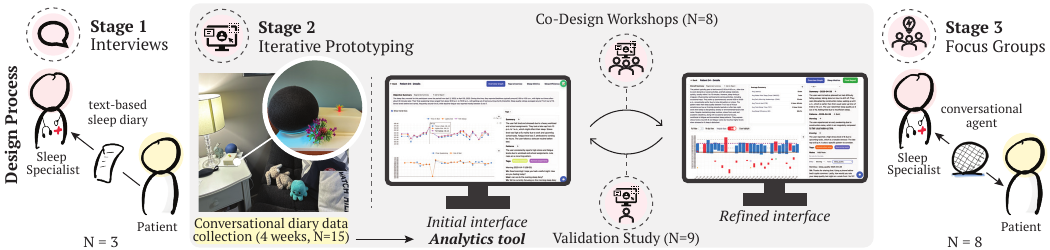}
  \caption{Our research process unfolded in three stages. In Stage 1, we interviewed behavioral sleep specialists to identify shortcomings of their current practice with text-based sleep diaries. In Stage 2, we designed a voice assistant sleep diary and conducted a 4-week deployment study with university students (proxy for patients) to collect data, followed by the iterative development of a \am{specialist-facing analytics tool} through initial prototyping, co-design workshops, prototype refinement, and a validation study. In Stage 3, we conducted focus groups with specialists to explore the broader potential of conversational agents in  BSM. }
  \Description{The figure depicts three stages of our research process from left to right. Stage 1, labeled ``Interviews,'' the picture illustrates the old process. The diagram shows the role labels ``Patient'' and “Sleep Specialist,” and the patient is handing a text-based sleep diary to the sleep specialist. Below this panel, the total number of participants is indicated as $N=3$ (three sleep specialists). Stage 2, labeled ``Iterative Prototyping,'' contains two steps. The first step shows an illustration of a desk with an Alexa device, labeled ``Conversational diary data collection (4 weeks, $t=15$).'' The second step, ``Analytics tool,'' includes two subpanels: the Initial interface on the left and the Refined interface on the right. Two arrows connect these subpanels: one from the Initial interface to the Refined interface labeled “Co-design Workshops ($N=8$),” and another from the Refined interface back to the Initial interface labeled ``Validation Study'' ($N=9$). Stage 3, labeled ``Focus Groups,'' shows a conversational agent pointing to both the Patient and the Sleep Specialist. Below this panel, the total number of participants is indicated as $N=8$ (eight focus groups).}
  \label{fig:process}
\end{figure}

\section{Research Overview}
We engaged sleep specialists in a three-stage design process (Fig. \ref{fig:process}): (1) \textit{Interviews} with specialists to characterize current diary use and to surface concrete shortcomings and unmet needs (Section \ref{sec:interviews}); (2) \textit{Iterative prototyping} of a conversational sleep diary and specialist-facing analytics tool through co-design workshops and a validation study, translating Stage 1 insights into refined prototypes (Section \ref{sec:interface-design}); and (3) \textit{Focus groups} with specialists, using the prototypes from Stage 2 as design probes to explore how conversational agents might complement diary-based workflows and surface potential use cases and boundaries in BSM practice. 

\noindent \textit{Participants. }
We engaged \am{thirteen} 
sleep specialists in our research process, recruited through snowball sampling from three groups: (1) clinical psychologists from Health System 1 ($n=3$), (2) \am{postdoctoral fellows from Health System 1 ($n=7$)}, and (3) postdoctoral fellows from different systems ($n=2$; one each from Health Systems 2 and 3). 
These participants allowed us to capture varied perspectives across career stages and institutions. Participant details are summarized in Table \ref{tab:participants}.
All study activities were approved by our institutional review board, and participants were compensated at a rate of \$15/hour.
\am{Across stages, specialist involvement varied due to scheduling constraints and the
distinct purpose of each activity. Stage 1 interview involved three specialists working across distinct professional roles and healthcare systems in BSM, which helped surface complementary views on how sleep diaries are currently interpreted and used. In Stage 2, eight (C1, C2, C3, F1, F2, F3, P1, and P2) specialists participated in workshops, either individually or in small groups, to iteratively refine the conversational diary and analytics tool. Five of these specialists (C1, P1, F1, F2, and F3) participated in the validation study along with four new participants (F4, F5, F6, and F7). Finally, in Stage 3, eight specialists (C1, C2, C3, F1, F2, F3, P1, and P2) took part in focus groups to discuss the potential role of CAs in BSM. }


\begin{table}[t]
\centering
\begin{minipage}{\textwidth}

\caption{\am{Demographics of participants. The activity column indicates research activities they participated in: interview (\textbf{\textcolor{orange}{I}}), Workshop and focus group (\textbf{\textcolor{cyan}{W}}), and validation study (\textbf{\textcolor{purple}{V}}). Column \textit{P} represents participant codes used throughout the manuscript.}} 
\label{tab:participants}
\centering
\begin{tabular}{l p{0.8cm} p{0.8cm} p{2.5cm} p{1cm} p{4.6cm} p{2cm}|p{1.2cm}}
\textbf{P} & \textbf{Gender} & \textbf{Age} & \textbf{Ethnicity} &
\textbf{Highest degree} & \textbf{Specialty---job title} &
\textbf{Years in clinic ---research} & \textbf{Activity} \\
\midrule[1.5pt]

\multicolumn{8}{l}{\textbf{Health System 1: Clinical Psychologists}} \\
\hline

C1 & F & 37 & Caucasian & PhD &
Behavioral Sleep Medicine---Clinical Psychologists \& Assistant Professor &
6---6 &
\textbf{\textcolor{orange}{I}} | \textbf{\textcolor{cyan}{W}} | \textbf{\textcolor{purple}{V}} \\

C2 & M & 57 & Hispanic/Latino/ Spanish & PhD &
Behavioral Sleep Medicine---Clinical Psychologists \& Associate Professor &
21---21 &
\textbf{\textcolor{cyan}{W}} \\

C3 & M & 58 & Caucasian & PhD &
Behavioral Sleep Medicine---Clinical Psychologists \& Professor &
27---27 &
\textbf{\textcolor{cyan}{W}} \\
\hline

\multicolumn{8}{l}{\textbf{Health System 1: Postdoctoral Fellows}} \\
\hline

F1 & F & 37 & Caucasian & PhD &
Behavioral Sleep Medicine---Fellow &
7---13 &
\textbf{\textcolor{orange}{I}} | \textbf{\textcolor{cyan}{W}} | \textbf{\textcolor{purple}{V}} \\

F2 & F & 39 & Caucasian & PhD &
Clinical Psychology---Fellow &
5.5---7 &
\textbf{\textcolor{cyan}{W}} | \textbf{\textcolor{purple}{V}} \\

F3 & F & 30 & Hispanic/Latino/ Spanish & PsyD &
Behavioral Sleep Medicine---Fellow &
6---0 &
\textbf{\textcolor{cyan}{W}} | \textbf{\textcolor{purple}{V}} \\

\am{F4} & F & 32 & Caucasian  & PhD &
Psychology---Fellow &
N.A. &
\textbf{\textcolor{purple}{V}} \\

\am{F5} & F & 33 & Caucasian & PhD &
Behavioral Medicine---Fellow &
1---7 &
\textbf{\textcolor{purple}{V}} \\

\am{F6} & F & 31 & Caucasian & PhD &
Behavioral Medicine---Fellow &
5---8 &
\textbf{\textcolor{purple}{V}} \\

\am{F7} & F & 32 & Asian & PhD &
Clinical Psychology, BSM---Fellow &
7---7 &
\textbf{\textcolor{purple}{V}} \\

\hline

\multicolumn{8}{l}{\textbf{Health System 2 and 3: Postdoctoral Fellows}} \\
\hline

P1 & F & 31 & Caucasian, Jewish & PhD &
Psychology (sleep)---Fellow &
8---10 &
\textbf{\textcolor{orange}{I}} | \textbf{\textcolor{cyan}{W}} \\

P2 & F & 41 & Hispanic/Latino/ Spanish & PsyD &
Psychology (sleep)---Fellow &
4---0 &
\textbf{\textcolor{cyan}{W}} | \textbf{\textcolor{purple}{V}} \\
\hline
\end{tabular}

\end{minipage}

\end{table}

\begin{table*}[t]
\begingroup
\centering
\caption{\am{Overview of specialists’ clinical---research split, percentage of times clinicians' use sleep diary in their practice, and other tools they use in BSM.}}
\label{tab:practice_overview}

\begin{tabular}{l l l l l p{8cm}}
\toprule
\textbf{P} & \multicolumn{2}{c}{\textbf{Week split}} 
& \multicolumn{2}{c}{\textbf{\am{Diary use}}} 
& \textbf{\am{Other tools used with or in place of sleep diaries}} \\
\cmidrule(lr){2-3} \cmidrule(lr){4-5}
& \textbf{Clinical} & \textbf{Research}
& \textbf{Clinical} & \textbf{Research} & \\
\midrule[1pt]

\textbf{C1} &
3.5 days & 1.5 days &
100\% & 70\% &
Actigraphy; Sleep EEG; PSG \\

\textbf{C2} &
1.5 days & 3.5 days &
100\% & 100\% &
N.A. \\

\textbf{C3} &
0.5 days & 4.5 days &
100\% & 100\% &
Wireless EEG; Actigraphy; Questionnaires \\

\textbf{F1} &
2 days & 3 days &
100\% & 50\% &
Actigraphy \\

\textbf{F2} &
0 days & 5 days &
N.A. & 50\% &
Actigraph; Sleep Profiler \\

\textbf{F3} &
5 days & 0 days &
100\% & N.A. &
N.A. \\

\textbf{F4} &
0 days & 5 days &
N.A. & 100\% &
N.A. \\

\textbf{F5} &
5 days & 0 days &
90\% & N.A.  &
N.A. \\

\textbf{F6} &
2 days & 3 days &
100\% & 100\% &
Questionnaires and clinical interviews collected through REDCap, such as ISI, MEQ,
SIS-D, FSS, GAD7, PHQ9, etc. \\

\textbf{F7} &
0 days & 5 days &
N.A. & 100\% &
N.A. \\

\textbf{P1} &
2 days & 3 days &
10\% & 0\% (currently) &
Insomnia Severity Index (ISI) \\

\textbf{P2} &
5 days & 0 days &
100\% & N.A. &
Actigraphy; Ambulatory EEG \\
\bottomrule

\end{tabular}
\endgroup
\end{table*}

\section{Stage 1: Interviews with Sleep Specialists to Identify Shortcomings of Traditional Sleep Diaries}
\label{sec:interviews}

We conducted one-on-one interviews with three sleep specialists---one from each group in Table \ref{tab:participants}: C1 (in-person), F1 (in-person), and P1 (online)---to better understand current practices in BSM, with a particular focus on the role of sleep diaries in behavioral therapy. We aimed to capture their experiences in clinical settings, surface shortcomings in current practices, and explore their vision for the ideal use of sleep diaries. \am{Each interview lasted about 30 minutes. Through three interviews spanning different specialties and health systems, we observed recurring themes and judged that additional interviews were unlikely to yield substantially new insights.}
\subsection{Data Analysis}

\am{All interviews were transcribed and analyzed using reflexive thematic analysis \cite{braun2023doing}}.  
The analysis focused on mapping the steps of BSM practice, identifying shortcomings and key challenges of the traditional sleep diaries, and deriving design requirements for a sleep diary tool to better support BSM. \am{Each interview transcript was analyzed independently by two members of the research team. The coders identified the shortcomings, key challenges, and design requirements mapped to the workflow of BSM. 
The coders then convened to extensively discuss and reach full agreement.}
Below we present our findings:

\subsection{Current Workflow and Shortcomings}

After referral from sleep specialist or primary care providers, the BSM workflow typically begins with a patient intake session, where the sleep diary is introduced. Patients then complete the diary between biweekly sessions. During the sessions, specialists review and use the entries to guide cognitive behavioral therapy (CBT) interventions (Fig. \ref{fig:interview-findings-workflow-shortcomings}).



\subsubsection{Patient intake session}
The intake session includes a diagnostic interview, a detailed patient history, and an introduction to CBT, with attention to whether the approach is suitable for the patient’s presenting disorder. As part of this session, specialists introduce the sleep diary and ask patients to complete it daily until the next visit.




\subsubsection{Recording sleep diary}
\label{sec:stage1:findings-recording-sleep-diary}
Patients record their sleep over the two-week period between visits. \am{As shown in Table \ref{tab:practice_overview}, sleep diaries are used extensively in both clinical and research settings for monitoring sleep behaviors in addition to other subjective questionnaires and objective tools (\eg actigraphy and EEG).} The diary recording process recurs across sessions and has following characteristics:
\begin{itemize} [leftmargin=*, nolistsep]
    \item \textit{Questions.} Questions in the diary are based on the consensus sleep diary \cite{carney2012consensus}.
    
    \noindent \textbf{Shortcomings.} This static diary offers little flexibility for different patients’ needs.  Memory erosion makes recall unreliable, and without contextual information about why certain nights were good or bad, the entries lack depth. 
    
    \item \textit{Format.} Administered typically in paper, digital (excel), or online forms (\eg REDCap). 
    
    \noindent \textbf{Shortcomings.} Participants highlighted several shortcomings in this process. 
    Patients often struggle with compliance and may retrospectively complete diaries. As F1 noted, \pquotes{``Even though they get prompted every day, they don't always fill it out every day \dots they might be retrospectively reporting compliance.''} Patients also sometimes misunderstand questions, leading to incorrect reporting. 
    Thus, filling diaries retrospectively outside of the suggested time windows and potential errors in filling diaries undermines data validity.
    
    \item \textit{Medium.} Patients either bring the diary to the visit or scan and email it to the specialist beforehand. 
    
    \noindent \textbf{Shortcomings.} Even when patients comply, diaries are often inaccessible to specialists before the session unless sent in advance. Reviewing them beforehand requires ``pajama time'' (unpaid preparation outside clinic hours).  
\end{itemize}

\begin{figure*}[t]
     \includegraphics[width=\textwidth]{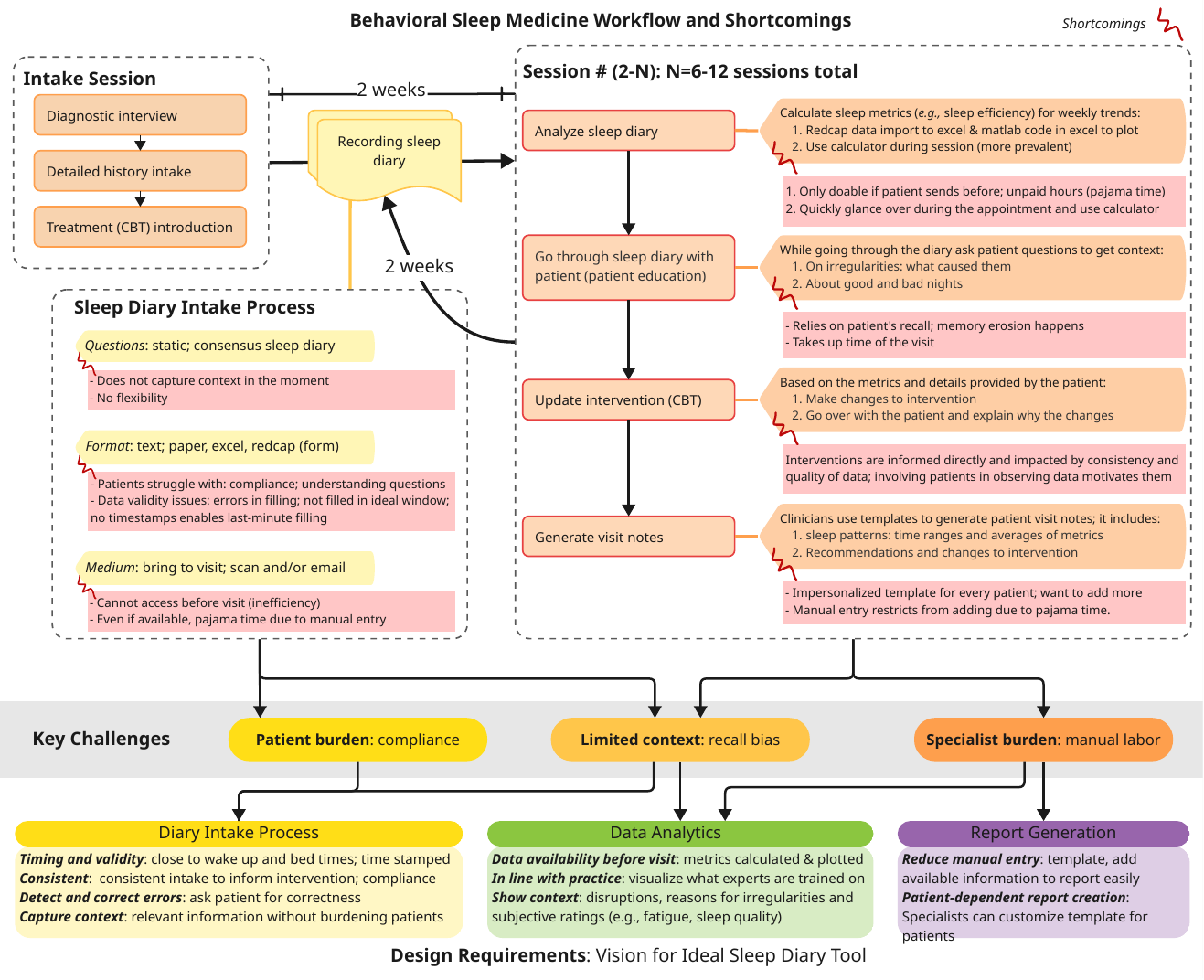}    
     \caption{Interview findings: Behavioral sleep medicine workflow and shortcomings in the use of traditional sleep diaries. We identified three key challenges in BSM: (1) patient burden, (2) lack of context reporting, and (3) specialist burden. Based on specialists' vision for an ideal sleep diary tool, we outline design requirements for our prototype (Stage 2). The \highlight{lightyellow}{part 1}, \highlight{lightgreen}{part 2} and \highlight{lightpurple}{part 3} of the prototype (stage 2) are color coded throughout the paper based on the design requirements.}
     \Description{This figure overall introduces the workflow, shortcomings, key challenges, and design requirements in behavioral sleep medicine. the figure is divided into three parts from top to bottom.  
    The first part is titled behavioral sleep medicine workflow and shortcomings, which is divided into two sides. the left side is intake session, which includes three subtitles: diagnostic interview, detailed history intake, and treatment (CBT-I) introduction. the right side is titled session # (2–n): n = 6–12 sessions total, which consists of four subsections: analyze sleep diary, go through sleep diary with patient (patient education), update intervention (cbt-i), and generate visit notes. for each subsection, there are additional explanations on the right along with shortcomings.  
    Analyze sleep diary explanation: calculate sleep metrics (e.g., sleep efficiency) for weekly trends:  
    1. Redcap data import to excel & matlab code in excel to plot.
    2. Use calculator during session (more prevalent).  
    Shortcomings:  
    1. Only doable if patient sends before; unpaid hours (pajama time).  
    2. Quickly glance over during the appointment and use calculator.  

    Go through sleep diary with patient (patient education) explanation: while going through the diary ask patient questions to get context:  
    1. On irregularities: what caused them.  
    2. About good and bad nights.
    Shortcomings:  
    1. Relies on patient's recall; memory erosion happens.  
    2. Takes up time of the visit.
    
    Update intervention (cbt-i) explanation: based on the metrics and details provided by the patient:  
    1. Make changes to intervention.
    2. Go over with the patient and explain why the changes  
    Shortcoming: interventions are informed directly and impacted by consistency and quality of data; involving patients in observing data motivates them .
    
    Generate visit notes explanation: clinicians use templates to generate patient visit notes; it includes:  
    1. Sleep patterns: time ranges and averages of metrics. 
    2. Recommendations and changes to intervention  
    Shortcomings: impersonalized template for every patient; want to add more and manual entry restricts from adding due to pajama time.  
    
    Between intake session and session #(2–n), there are three connecting lines. One line without arrows in the middle is labeled 2 weeks, indicating the frequency of occurrence. below it, there is another line from intake session pointing to session # (2–n) with a folder icon in the middle labeled recording sleep diary. At the same time, session # (2–n) also has a line pointing back to recording sleep diary with 2 weeks written in the middle, showing the overall workflow direction. Below recording sleep diary, there is a dashed line connected to an explanation titled sleep diary intake process:  
    Questions: static, consensus sleep diary  
    Shortcomings: does not capture context in the moment and no flexibility. 
    Format: text, paper, excel, redcap (form)  
    Shortcomings: patients struggle with compliance; understanding questions and data validity issues: errors in filling; not filled in ideal window; no timestamps enables last-minute filling.  
    Medium: bring to visit; scan and/or email  
    Shortcomings: cannot access before visit (inefficiency) and even if available, pajama time due to manual entry.  
    
    The second part is titled key challenges, which is divided into three items from left to right: patient burden: compliance, limited context: recall bias, and specialist burden: manual labor. Among them, patient burden has a line pointing from the sleep diary intake process in the first part. Limited context has a line pointing from both the sleep diary intake process and session # (2–n) in the first part. Specialist burden has a line pointing from session # (2–n) in the first part.  
    
    The third part is titled design requirements: vision for ideal sleep diary tool, which is divided into three sections from left to right: diary intake process, data analytics, and report generation.  
    
    Diary intake process includes:  
    - Timing and validity: close to wake up and bed times; time stamped.  
    - Consistent: consistent intake to inform intervention; compliance.  
    - Detect and correct errors: ask patient for correctness  
    - Capture context: relevant information without burdening patients.  
    It has a line pointing from patient burden in the second part and another line from limited context in the second part.  
    
    Data analytics includes:  
    - Data availability before visit: metrics calculated \& plotted.  
    - In line with practice: visualize what experts are trained on.  
    - Show context: disruptions, reasons for irregularities and subjective ratings (e.g., fatigue, sleep quality)  
    it has a line pointing from specialist burden in the second part and another line from limited context in the second part.  
    
    Report generation includes:  
    - Reduce manual entry: template, add available information to report easily.  
    - Patient-dependent report creation: specialists can customize template for patients.  
    It has a line pointing from specialist burden in the second part.
    }
    \label{fig:interview-findings-workflow-shortcomings}
\end{figure*}

\subsubsection{Biweekly patient-specialist sessions}
\label{sec:stage1-findings-biweekly sessions}
The biweekly sessions are structured as follows: 

\begin{enumerate} [leftmargin=*, nolistsep]
    \item \textit{Analyze sleep diary.} Specialists calculate sleep metrics such as time in bed (TIB), total sleep time (TST), and sleep efficiency (SE), often using calculators or Excel scripts.

   \noindent \textbf{Shortcomings.} Most specialists rely on quick calculations during the session when using paper-based diaries (2/3 participants used paper diaries), which makes the process inefficient—especially given that sleep interventions are highly dependent on these metrics. In the case of online forms such as REDCap, the process still requires substantial manual labor, C1: \pquotes{``It's a clunky process, so we have to download it from REDCap, and then we have to input it into this master Excel copy, and then it will calculate everything for us. So somebody has to schedule the REDCap diaries, make sure that they're getting completed, download the Excel, put it into this\dots''}

   \item \textit{Go through sleep diary with patient.} Specialists review diary data with patients, asking about irregularities, stressors, and coping techniques to get context. 

    \noindent \textbf{Shortcomings.} This process is time-consuming and relies heavily on patients’ written comments in the diary and their ability to recall. Inconsistency in recording these comments and memory erosion are major challenges, C1: \pquotes{``Most people don't fill out the comments with too much detail. And then you're sitting down with them and you're like, Wow, this was a really bad night\dots What happened? And they're like, I have no idea.''} 
    
    \item \textit{Update intervention.}
    Based on quantitative metrics (from sleep diary data) and qualitative aspects (from conversations with patients), specialists update interventions as needed and explain the changes and their rationale to patients.

    \noindent \textbf{Shortcomings.} Because interventions are directly informed by diary data, consistency and quality are crucial. However, they are often compromised by low compliance, retrospective completion (recall), and potential errors. 

    \item \textit{Generate visit notes.}
    Towards the end of the visit, specialists generate reports accessible to patients. These rely on templates that summarize: (1) sleep patterns, including time ranges and averages of metrics, and (2) interventions.
    
    \noindent \textbf{Shortcomings.} Templates are impersonal. Although specialists want to personalize them, manual entry makes customization difficult and adds to pajama time, C1: \pquotes{``There's probably a lot of information I'd like to include, but for time reasons, I have everything templated, and I try to do it really quickly.''}
    
\end{enumerate}

\subsection{Key Challenges in Current Practice}
We identified three main challenges (Fig. \ref{fig:interview-findings-workflow-shortcomings}):  
(1) patient burden, which leads to compliance challenges and hinders consistent, accurate, and valid diary recordings (Section \ref{sec:stage1:findings-recording-sleep-diary});  
(2) limited context captured in-the-moment causes recall bias (Section \ref{sec:stage1:findings-recording-sleep-diary} and \ref{sec:stage1-findings-biweekly sessions}); and  
(3) clinician burden due to limited data availability and insufficient tools for data analysis (Section \ref{sec:stage1-findings-biweekly sessions}). Next, we identify the design requirements to address these challenges. 

\subsection{Vision for Ideal Sleep Diary Tool $\rightarrow$ Design Requirements}
\label{sec:stage1-design-requirements}
When asked about their vision for an ideal tool to address the discussed key challenges, specialists articulated requirements across three areas: diary intake, data analytics, and report generation (Fig. \ref{fig:interview-findings-workflow-shortcomings}).

\subsubsection{Requirements for sleep diary intake process}
To address challenges of \textit{patient burden} and \textit{limited context}, the ideal tool should meet these requirements:
\begin{itemize} [leftmargin=*, nolistsep]
    \item \textit{\highlight{DR_yellow}{R1}Timing and validity.} Entries should be completed close to sleep and wake times and be time-stamped, P1: \pquotes{``My instructions are always fill it within, ideally, 30 minutes of waking up, maximum 60.''} 

    \item \textit{\highlight{DR_yellow}{R2}Consistency. } Entries should be completed consistently to support effective interventions, especially for clinical research trials, C1: \pquotes{``Especially to make sure we have timestamps.''}. 

    \item \textit{\highlight{DR_yellow}{R3}Detect and correct errors. } The tool should detect unusual values and confirm with patients during intake, F1: \pquotes{``Any unusual values, some people just make mistakes. So if they [tool] could be like, did you really mean to say that you were going to bed at 11 a.m.?''}

    \item \textit{\highlight{DR_yellow}{R4}Capture context.} The tool should capture context---such as reasons for irregularities, environmental conditions, stress, and fatigue---without overburdening patients, C1:  \pquotes{``If it [tool] prompted them to follow up a little bit more information.''} and \pquotes{``capturing as much information as possible about the context of the night without burdening them.''}

\end{itemize}

\subsubsection{Requirements for data analytics}
To address challenges of \textit{specialist burden} and \textit{limited context} in available data, the ideal tool should meet these requirements:

\begin{itemize} [leftmargin=*, nolistsep]
    \item \textit{\highlight{DR_green}{R5}Data availability before visit.} Data should be automatically processed and made available to the specialist before the session with the patient (C1: \pquotes{``Being able to see it before the patient comes.''}). 

    \item \textit{\highlight{DR_green}{R6}In line with practice.} Visualizations should at minimum include the data, metrics, and plots specialists are trained on. Thus, the tool should incorporate the quantitative metrics commonly relied upon in practice. 

    \item \textit{\highlight{DR_green}{R7}Show context.} The tool should surface contextual factors---such as disruptions or qualitative explanations for patients’ subjective ratings (e.g., sleep quality)---in addition to the quantitative metrics. 
\end{itemize}

\subsubsection{Requirements for report generation}

To address \textit{specialist burden},  ideal tool should meet these requirements:

\begin{itemize} [leftmargin=*, nolistsep]
    \item \textit{\highlight{DR_purple}{\textcolor{white}{R8}}Reduce manual entry.} Reports should minimize manual effort and automatically populate the basic information specialists typically add, \eg \pquotes{``a summary report of data that would be helpful''} (F1) and \pquotes{``I would include those key [quantitative] metrics''} (P1). 

    \item \textit{\highlight{DR_purple}{\textcolor{white}{R9}}Patient-dependent report creation.} Reports should allow specialists to customize templates for each patient by selecting relevant content, since patients can have different preferences or comfort level around seeing their sleep data, P1: \pquotes{``Some patients get a printout of it and they love it, versus people who are a little less data oriented\dots don't care as much.''}  

\end{itemize}


\section{Stage 2: Iterative Prototyping of Sleep Diary Tool and Specialist-Facing Interface}
\label{sec:interface-design}

In Stage 2, we iteratively prototyped a conversational sleep diary and a data \am{analytics tool} in collaboration with specialists. We first developed a voice assistant diary to collect proxy patient data, then developed an interface to help specialists interpret it.
\am{We first developed a voice assistant diary to collect proxy-patient data from 15 university students\footnote{The data collection was approved by Institutional Review Board and participants were compensated.}, which allowed us to generate realistic conversational diary entries at scale without introducing clinical risk or patient burden at an early stage of the design process. Our goal in this deployment was not to generalize findings to clinical populations, but to obtain  real examples of diary conversations that could be used to observe how specialists interact with and interpret conversational data.}
Feedback from co-design workshops helped us refine the interface, which we subsequently evaluated for usability and alignment with clinical practice. 

\subsection{Initial prototype}

The prototype consisted of three parts: 
(1) a conversational diary, (2) \am{an analytics tool}, and (3) report generation, each incorporating requirements identified during interviews (Section \ref{sec:stage1-design-requirements}).

\subsubsection{Initial Prototype  \highlight{lightyellow}{Part 1}: Conversational VA Sleep Diary}
In this part, we designed a voice-based sleep diary tool to capture important aspects of sleep. To enable natural conversations, we employed LLMs \cite{mahmood2025user, mahmood2025voice, yang2024talk2care}.

\noindent \textbf{LLM-powered VA diary tool.} To support a natural, conversational way of completing sleep diaries, we developed a voice assistant that runs on Amazon Alexa and is powered by ChatGPT similar to prior work on LLM-powered Alexa \cite{mahmood2025user}. The VA proactively reminds users in the morning and evening to finish their diaries, prompting them until they are finished or the time to finish the diary has passed. At the same time, users can also activate it themselves by simply saying phrases like ``Alexa, sleep diary.'' Once active, the VA guides users through a set of clinically established sleep diary questions. 
In line with the  \highlight{DR_yellow}{R1}\textit{timing and validity}, and \highlight{DR_yellow}{R2}\textit{consistency} requirement, the VA adapts to each user’s schedule, ensures diaries are completed close to relevant sleep/wake times, and records timestamped conversations for later review by specialists.

\begin{figure*}[t!]
     \includegraphics[width=\textwidth]{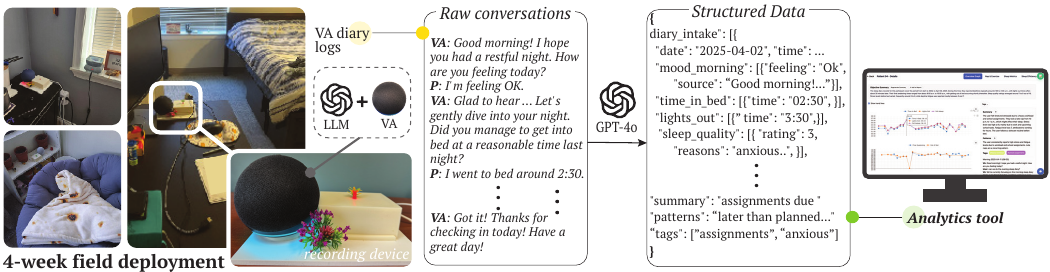}    
     \caption{Stage 2: LLM-powered VA sleep diary four-week data collection study with 15 university students as proxy patients. The raw diary conversations are then converted to structured JSON objects for use in the analytics tool.}
     \Description{This figure explains the overall process of the llm-powered VA sleep diary four-week data collection study with 15 university students as proxy patients. On the left side, there are three photos showing the real experimental participants' recording device placement, along with a zoomed-in image of the recording device. On the right side of this part, it is explained as a combination of llm+va. This section has an overall title 4-week field deployment. To its right is a title VA diary logs, which points to a raw conversations section. this section contains a sample dialogue as follows:  

    VA: Good morning! I hope you had a restful night. How are you feeling today?  
    P: I’m feeling OK.  
    VA: Glad to hear ... let’s gently dive into your night. Did you manage to get into bed at a reasonable time last night?  
    P: I went to bed around 2:30.  
    ...  
    VA: Got it! Thanks for checking in today! Have a great day!  

    To the right of raw conversations, there is another part structured data, connected by an arrow from raw conversations to structured data labeled gpt-4o, indicating the processing direction. Structured data shows a sample format of the data as follows:  
    
    {diary_intake=[{  
      "date": "2025-04-02", "time": ...,  
      "mood_morning": {"feeling": "ok",  
                       "source": "good morning..."},  
      "time_in_bed": {"time": "02:30"},  
      "lights_out": {"time": "3:50"},  
      "sleep_quality": {"rating": 3,  
                        "reasons": "anxious..."}  
    }],  
    ...  
    summary: "assignments due"  
    patterns: "later than planned..."  
    tags: ["assignments", "anxious"]}
    
    To its right, there is a virtual computer screen image, labeled at the bottom as analytics tool.  
    }
    \label{fig:stage2-initial-prototype-VA-diary} 
\end{figure*}

\noindent \textbf{Sleep diary questions.}
We designed the questions based on the consensus sleep diary provided by specialists, then adapted them for our target population of university students in consultation with three specialists (C1, C2, C3). Morning diaries focused on sleep timing, awakenings, and quality, while evening diaries captured information such as naps, caffeine and alcohol intake, stress, fatigue, and bedtime routines. The conversational nature of the VA allowed us to elicit context for users’ perceptions of sleep through occasional follow-up questions. In line with the \highlight{DR_yellow}{R4}\textit{capture context} requirement, the VA probed for reasons behind subjective ratings (\eg sleep quality, stress, or fatigue) and asked to expand on other relevant aspects of their sleep experiences. In line with the \highlight{DR_yellow}{R3}\textit{detect and correct errors} requirement, the VA was designed to flag potential inaccuracies and confirm them with the user. It also repeated captured numbers, allowing users to catch and correct errors when needed.

\noindent \textbf{Participants and data collection.}
We deployed the VA diary with 15 university students (8 female, 7 male; undergraduate and graduate), aged 20–32 ($M=24.27$, $SD=3.95$), over a four-week period (Fig. \ref{fig:stage2-initial-prototype-VA-diary}). We collected diary data through conversation logs and audio data. We collected about 80 hours of data. Completion rates ranged from 78–100\% ($M=92.00\%$, $SD=7.82\%$).

\subsubsection{Initial Prototype  \highlight{lightgreen}{Part 2}: \am{Specialist-Facing Analytics Tool---From Conversations to Insights}}

\am{This part focused on designing an interface that presents the collected diaries alongside standard sleep metrics, augmented with conversational context that specialists can readily interpret.
}

\noindent \textbf{Diary logs-to-JSON conversion.} 
To meet the \highlight{DR_green}{R5}\textit{data availability before visit} requirement, unstructured diary logs were gathered and parsed into structured JSON objects automatically---using zero-shot prompting with GPT-4o, while preserving original conversations\footnote{All prompts for the initial and refined prototype are available here: \url{https://tinyurl.com/2vvc95cw}.} (Fig. \ref{fig:stage2-initial-prototype-VA-diary}). 
\am{To evaluate the reliability of the LLM-based diary-to-JSON conversion, we manually checked approximately 20\% of the dataset by randomly sampling 10 diaries from each participant (five morning and five evening entries), resulting in 150 diary entries across 15 participants. Within this subset, two coders first annotated 30 diaries (20\% of subset) to establish intercoder reliability. In our coding scheme, completeness indicates whether all required information from the diary appears in the JSON output, while accuracy reflects whether the extracted value matches the conversation (details in Appendix \ref{app:json}). The percentage agreements for completeness and accuracy are $99.76\%$ and $98.78\%$, respectively. Across this subset, 99.75\% of diary elements were captured---of which 99.44\% were fully complete and 0.31\% partially complete---and 97.71\% were accurate, indicating that the JSON outputs were sufficiently reliable for the sake of this work.}
The structured data \am{comprising of clinically important sleep behaviors} was fed to the specialist interface. The interface (Fig. \ref{fig:stage2-initial-prototype}) comprised five panels---\textit{Summary}, \textit{Visualization}, \textit{Conversation}, \textit{Chat Bubble}, and \textit{Report}---and integrated both quantitative and qualitative insights:

\begin{figure*}[t]
     \includegraphics[width=\textwidth]{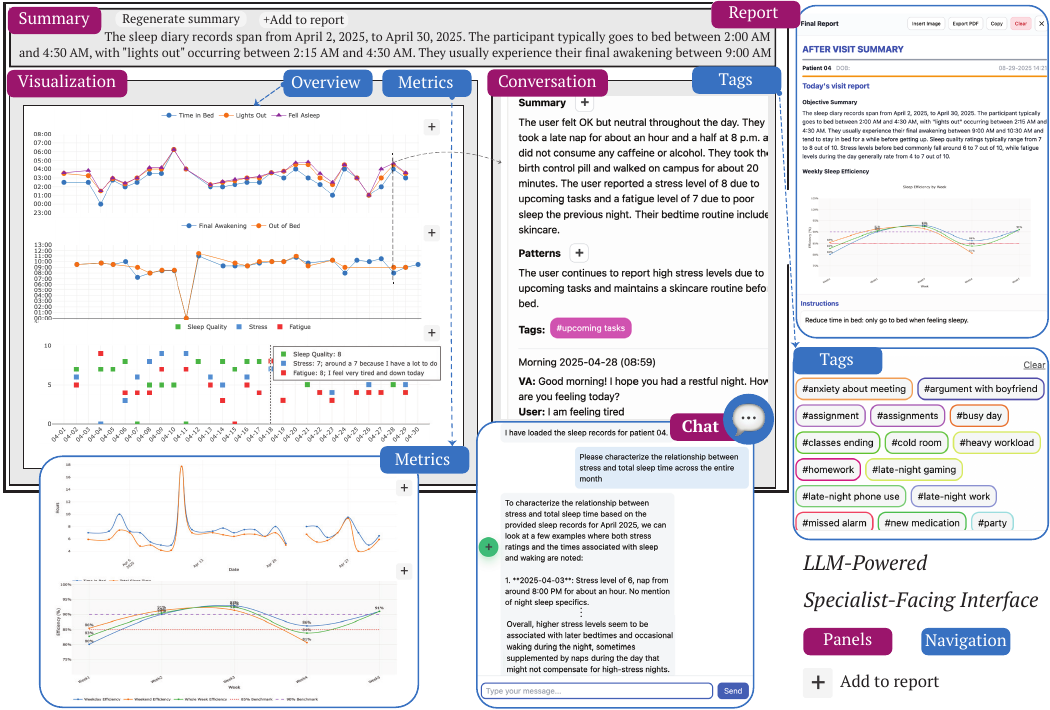}    
     \caption{Stage 2: Initial prototype for LLM-powered specialist-facing interface (part 2 and 3). Qualitative and quantitative insights along with report generations are implemented in the five panels: \textit{Summary}, \textit{Visualization}, \textit{Conversation}, \textit{Chat Bubble}, and \textit{Report}.}
     \Description{This figure shows the initial prototype for llm-powered specialist-facing interface, which is divided into five parts to display the interface. There are also two types of titles: pink titles represent panels, and blue titles represent navigation. At the top left is a complete interface, which is divided into three parts: summary (pink title), visualization (pink title), and conversation (pink title). The summary (pink title) section of the interface has two buttons at the top, namely ``regenerate summary'' and ``+add to report''. Below is a paragraph of text: ``the sleep diary records span from april 2, 2025, to april 30, 2025. The participant typically goes to bed between 2:00 am and 4:30 am, with 'lights out' occurring between 2:15 am and 4:30 am. They usually experience their final awakening between 9:00 am''.  

     On the bottom left of the summary (pink title) section is the visualization (pink title) section, which contains a line chart and a scatter plot, and an overview (blue title) pointing to these two charts. On the right side of the visualization (pink title) section is the conversation (pink title) section. in the interface section, the text includes:  
     Summary: The user felt OK but neutral throughout the day. They took a late nap for about an hour and a half at 8 p.m., did not consume any caffeine or alcohol. They took their birth control pill and walked on campus for about $20$ minutes. The user reported a stress level of $8$ due to upcoming tasks and a fatigue level of $7$ due to poor sleep the previous night. Their bedtime routine includes skincare.  
     Patterns: The user continues to report high stress levels due to upcoming tasks and maintains a skincare routine before bed.
     Tags: #upcoming tasks  
     Morning 2025-04-28 (08:59)  
     VA: good morning! I hope you had a restful night. how are you feeling today?  
     User: I am feeling tired.  
     
     To the right of this part, there is a smaller section showing the report (pink title) section, and below it another section tags (blue title). This part displays sample tags: #anxiety about meeting, #argument with boyfriend, #assignment, #assignments, #busy day, #classes ending, #cold room, #heavy workload, #homework, #late-night gaming, #late-night phone use, #late-night work, #missed alarm, #new medication, #party.  

    To the left of this, there are also two sections Metrics (blue title) and chat (pink title). Metrics (blue title) displays two line charts. Chat (pink title) displays a short conversation.  
    }
    \label{fig:stage2-initial-prototype} 
\end{figure*}

\noindent \textbf{1. Quantitative insights.}
To meet the \highlight{DR_green}{R6}\textit{in line with current practice} requirement, the interface provided two types of plots in the \textit{Visualization Panel}:

\begin{enumerate} [ nolistsep]
    \item \textbf{Overview graph:} The \textit{Overview} tab displayed daily raw values (\eg time got in bed, lights out, sleep onset, final awakening, out-of-bed times, naps, and exercise) alongside subjective ratings of sleep quality, stress, fatigue on a scale of 1 to 10. Clicking a data point revealed the original conversation, an LLM-generated summary, and patterns for the day in the \textit{Conversation Panel}.

    \item \textbf{Sleep metrics:} Automatically calculated metrics (\eg time in bed (TIB) vs. total sleep time (TST), weekly sleep efficiency averages) were plotted. These are the same metrics specialists typically use to guide interventions. 

\end{enumerate}

  All plots were interactive, allowing specialists to select or deselect elements to reduce visual clutter and examine trends either in combination or in isolation.

\noindent \textbf{2. Qualitative insights.}
In line with the \highlight{DR_green}{R7}\textit{show context} requirement, the interface incorporated qualitative features derived from the conversational data:

\begin{enumerate} [ nolistsep]
    \item \textbf{Reasons for subjective ratings.} Hovering over subjective rating data points (\textit{Overview} tab) displayed patient-provided explanations for stress, fatigue, or sleep quality extracted during diary logs-to-JSON conversion. 

    \item \textbf{Tags for diary entries:} During diary logs-to-JSON conversion, GPT generated reason-focused tags (\eg stressors, facilitators, unusual events). Tags were shown in the \textit{Conversation Panel}; clicking a tag highlighted its occurrences in the overview graph. A dropdown menu above the \textit{Conversation Panel} listed all patient tags.

    \item \textbf{Conversational querying:} The \textit{Chat Bubble} allowed specialists to pose free-form questions to an LLM-powered agent, which had access to the structured JSON---containing the original conversations as well. LLM-powered agent could highlight relevant data points---mentioned in its response---in the \textit{Overview} graph, allowing specialists to refer back to original conversations. 

\end{enumerate}

\subsubsection{Initial Prototype  \highlight{lightpurple}{Part 3}: Report Generation for Patient Communication}

To support the requirements of \highlight{DR_purple}{\textcolor{white}{R8}}\textit{reducing manual entry} and \highlight{DR_purple}{\textcolor{white}{R9}}\textit{patient-dependent report creation}, we implemented a flexible report-building feature. Specialists could add any element of the interface---plots from \textit{Visualization} panel, daily summaries or patterns from the \textit{Conversation Panel}, or any selected text from the \textit{Chat Bubble}---directly into the report. An AI-generated four-week summary (\textit{Summary Panel}) highlighted overall patterns and could be included with a single click.
Additionally, the report was fully editable: specialists could add, remove, or rearrange content as needed. Once finalized, it could be copied into other systems or exported as a PDF for sharing with patients.

\subsection{Co-design Workshops with Sleep Specialists}

The co-design workshops aimed to gather specialists’ feedback on the conversational diary prototype, with a focus on the data analytics interface (part 2) and report generation features (part 3). We conducted five sessions with eight participants (Table \ref{tab:participants}): W1 (C1), W2 (C2, C3), W3 (F1, F2, F3), W4 (P1), and W5 (P2). Workshops W1–W3 were conducted in person, while W4 and W5 were conducted remotely via Zoom. 
\am{Workshops varied in participation size due to specialist availability. Although some sessions (W1, W4, W5) involved a single participant, all workshops followed the same procedure and were intended to gather prototype-specific feedback. Each participant interacted with the prototype on their personal laptops to ensure independent exploration. In the group sessions (W2 and W3), collaborative discussion among participants was encouraged. All sessions lasted between 30--60 minutes.} 

\subsubsection{Procedure}
Each workshop followed a two-step process. First, participants were introduced to the conversational VA used for diary intake (part 1), and audio recordings of one morning and one evening diary were played to illustrate the type of conversational data collected. Next, participants interacted with the high-fidelity interface to explore features for data analytics (part 2) and report generation (part 3). During this stage, they were encouraged to ``think aloud'' and share feedback on how the tool could be improved to better fit clinical practice. 

\subsubsection{Data Analysis} 
We conducted \am{reflexive thematic analysis \cite{braun2023doing}} on the transcribed workshop audio and participants’ interactions with the interface to identify feedback and improvements suggested for data analytics and report generation. \am{Two authors first independently coded all transcripts and interaction data for suggested improvements and features. These improvements and features were then organized into themes and sub-themes. The two coders then convened to extensively discuss and reach full consensus on the coding.} The resulting themes were: (1) data analytics, (2) AI-assisted analysis, and (3) report generation. Fig.~\ref{fig:stage2-workshop-findings} summarizes these findings; the highlights below correspond to the color codes in the figure. We now detail the sub-themes and proposed features and improvements: 

\begin{figure*}[t]
     \includegraphics[width=\textwidth]{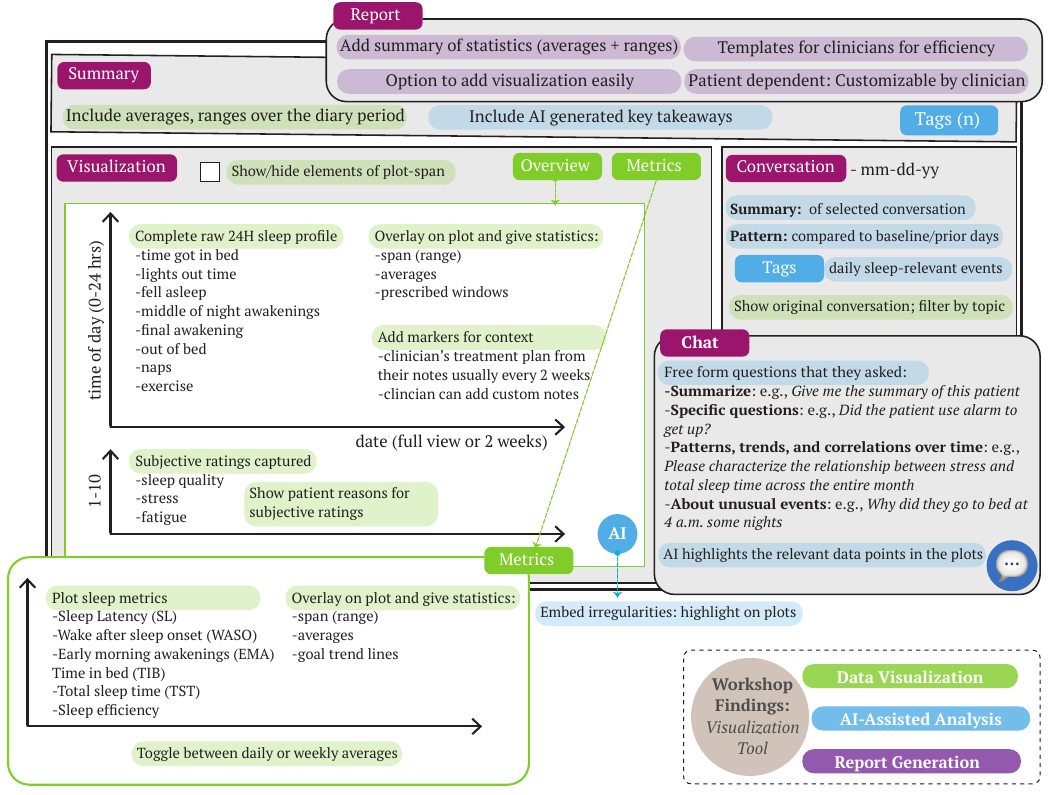}    
     \caption{Stage 2: Co-design workshop findings for specialist-facing interface. To refine the interface, various features under three themes---data visualization, AI-assisted analysis, and report generation---are implemented in the five panels: \textit{Summary}, \textit{Visualization}, \textit{Conversation}, \textit{Chat Bubble}, and \textit{Report}.}
     \Description{This figure shows the planned modifications to the interface after the co-design workshop. In the bottom-right corner of the figure, there is a legend that categorizes various features under three themes: data visualization (green), AI-assisted analysis (blue), and report generation (purple).  
     In the report section, the items include: add summary of statistics (averages + ranges) (purple), templates for clinicians for efficiency (purple), option to add visualization easily (purple), and patient dependent: customizable by clinician (purple).  
     In the summary section, the items include: include averages, ranges over the diary period (green), include ai generated key takeaways (blue), and tags (n) (blue).  
     In the visualization section at the lower left of summary, there is a checkbox show/hide elements of plot-span (green). there are also two green-titled subsections: overview and metrics.  
     For overview, it contains two coordinate axes. The first coordinate axis has x-axis as date (full view or 2 weeks) and y-axis as time of day (0-24 hrs). Within this axis, there are three subsections: complete raw 24h sleep profile (green), overlay on plot and give statistics (green), and add markers for context (green). 
     - complete raw 24h sleep profile includes: time got in   bed, lights out time, fell asleep, middle of night      awakenings, final awakening, out of bed, naps, and       exercise.  
     - overlay on plot and give statistics includes: span (range), averages, and prescribed windows.  
     - add markers for context includes: clinician’s treatment plan from their notes usually every 2 weeks, and clinician can add custom notes.  
     The second coordinate axis has x-axis as empty and y-axis as 1-10. Within this axis, there are two subsections: subjective ratings captured (green) and show patient reasons for subjective ratings (green).  
     - subjective ratings captured includes: sleep quality, stress, and fatigue.  
     For metrics, it also contains a coordinate axis without labeled x-axis and y-axis. Below it there is a title toggle between daily or weekly averages. Inside this section, there are two subsections: plot sleep metrics (green) and overlay on plot and give statistics (green). 
     - plot sleep metrics includes: sleep latency (sl), wake after sleep onset (waso), early morning awakenings (ema), time in bed (tib), total sleep time (tst), and sleep efficiency.  
     - overlay on plot and give statistics includes: span (range), averages, and goal trend lines.  
     At the bottom-right of visualization, there is a blue circular button labeled ai, explained as embed irregularities: highlight on plots (blue).  
     To the right of the visualization section, there is the conversation– mm-dd-yy section. This section contains: summary: of selected conversation (blue), pattern: compared to baseline/prior days (blue), tags/ daily sleep-relevant events (blue), and show original conversation; filter by topic (green).  
     Below the conversation section is the chat section, which contains two subsections: free form questions that they asked (blue) and ai highlights the relevant data points in the plots (blue).  
     - free form questions that they asked includes: summarize: e.g., give me the summary of this patient; specific questions: e.g., did the patient use alarm to get up?; patterns, trends, and correlations over time: e.g., please characterize the relationship between stress and total sleep time across the entire month; and about unusual events: e.g., why did they go to bed at 4 a.m. some nights?  
     }
    \label{fig:stage2-workshop-findings} 
\end{figure*}

\subsubsection{Findings:  \highlight{lightgreen}{Data Visualization}}
We identified three sub-themes related to data visualization (Fig. \ref{fig:stage2-workshop-findings}): 

\noindent \textbf{Match and integrate into current practice.}

Specialists appreciated that the visualizations were grounded in their practice but wanted all data presented more compactly alongside information from treatment plans. For the \textit{Overview} graph—showing raw data such as lights-out time, final awakening, and naps—participants requested a full 24-hour sleep profile to capture atypical patterns (F2: \pquotes{``a 24 hour scale, like the people\dots who sleep atypical times.''}). They also highlighted the value of displaying prescribed windows and treatment goals directly on the plots for quick visual comparison (F1: \pquotes{``It would be nice if I could add a trend line. We give people prescribed bedtimes and wake times.''}). In addition, they valued summary statistics such as ranges and averages in the \textit{Summary Panel}, but emphasized the need to view two weeks at a time to align with the biweekly rhythm of clinical sessions.


\noindent \textbf{Overlay context.}
Participants emphasized the importance of contextualizing the data. They appreciated visualizations embedded with with patients’ comments, such as reasons for subjective ratings (on hover)  or other influencing factors (\textit{Tags}). Specialists additionally wanted to include their inputs, including treatment plans and observations, C3: \pquotes{``It would be good also to be able to put in markers so if something important happened, one treatment started, one treatment ended.''}

\noindent \textbf{Quick access to original conversations.}
Specialists valued being able to open the \textit{Conversation Panel} by clicking on a data point, giving them direct access to the original transcript for validating plotted data. They also requested filtering options to quickly locate conversations by topic (\eg stress, fatigue, wake time, sleep environment), since full transcripts can be lengthy and difficult to parse.

Together, these findings emphasize specialists’ desire for visualizations that both align with existing workflows and enrich them with contextual patient data. 

\subsubsection{Findings: \highlight{lightblue}{AI-Assisted Analysis}}
We identified four sub-themes related to AI-assisted analysis (Fig. \ref{fig:stage2-workshop-findings}):

\noindent \textbf{Summarization.}  They appreciated the summary feature at two levels: (1) across the full diary period, to highlight key takeaways in the \textit{Summary Panel}, and (2) for each diary entry, in the \textit{Conversation Panel} as a quick daily snapshot without reading the full transcript. They also valued the ``patterns'' feature, which highlights how the current entry differs from prior days. Participants suggested using early data (\eg week 1) as a baseline to better identify deviations. Combined with topic-based filtering, this enables a top-down approach for quickly locating relevant information. 

\noindent \textbf{Irregularities embedded in visualization.} Specialists emphasized the value of seamless AI integration into visualizations to surface trends relevant to the treatment. They envisioned irregularities being flagged automatically with a single click. Irregularities could be detected based on (1) established clinical guidelines (\eg WASO > 30 minutes), and (2) deviations from a patient’s usual or prescribed schedule (\eg unusually late bedtime). They also wanted contextual explanations overlaid directly on the visualization, extracted from conversations (\eg highlighting ``headache'' when a late bedtime occurs), P1: \pquotes{``This was unusual for them, because they had a headache. So to be able to see that in the comment up here rather than scrolling down through the whole transcript, might be a nice highlight.''}.

\noindent \textbf{Tag sleep-relevant events.} Specialists valued the \textit{Tags} feature for tracking recurring stressors and monitoring their change over treatment, C1: \pquotes{``hopefully see over time that they can start to change in terms of what sticks out.''}. They suggested categorizing tags into (1) internal disruptions (\eg schoolwork, travel), (2) external disruptions (\eg construction noise), and (3) salient positive/negative events (\eg graduation, interviews), P1: \pquotes{``maybe collapse across categories of an internal disruption or an external disruption''}). Participants also emphasized showing tag frequencies to quickly indicate repeated factors. 

\noindent \textbf{Conversational querying. }
Specialists generally appreciated the \textit{Chat Bubble} as a flexible way to explore patient data. From their interaction logs, we observed three types of questions they asked: (1) requests for overviews or summaries (\eg F1: \pquotes{``Give me summary of this patient.''; F2: \pquotes{``What are some of the sleep irregularities?''}}. (2) questions about patterns and correlations among metrics over time (\eg C2: \pquotes{``Please characterize the relationship between stress and total sleep time across the entire month.''}. (3) targeted questions when noticing anomalies in visualizations (\eg C1 asked \pquotes{``Why did they go to bed at 4:00 am some nights?''}. 

Together, these findings highlight specialists’ desire for AI tools that not only summarize and flag irregularities, but also contextualize them for patient care and enable deeper and more efficient engagement with data.

\subsubsection{Findings: \highlight{lightpurple}{Report Generation}}
\label{sec:stage2-workshop-findings-reporting}
We identified three sub-themes related to report generation (Fig. \ref{fig:stage2-workshop-findings}):

\noindent \textbf{Bare-bones structure.}
Participants noted that reports should at minimum include an overview of sleep statistics (averages and ranges) and key insights generated by the LLM in the \textit{Summary Panel}. They also expected specialist notes and planned interventions to be incorporated. To streamline this process, specialists suggested creating a basic template that the system could auto-populate for efficiency.

\noindent \textbf{Customization by specialist.}
Beyond a base template, participants emphasized the need for flexibility to tailor reports to individual patients, C3: \pquotes{``We could come up with some sort of summary, but then being able to tailor that. Once you meet with them, you might want to add stuff, take stuff away or whatever.''}

\noindent \textbf{Adding visualizations. }
Specialists expressed caution about daily plots, noting that some patients may become overly focused on night-to-night variability. Instead, they recommended emphasizing weekly summaries, which provide a more stable picture of progress. 
They also emphasized the value of goal-oriented trend lines to contextualize progress and reduce misinterpretation, C1: \pquotes{``[Showing] here's what your goal is and here's where you are relative to that goal\dots I [patient] stayed within the lines.''}).

Together, these findings highlight specialists’ desire for reports that balance efficiency through auto-populated templates, flexibility through customization, and patient-centered communication through carefully chosen visualizations and goal-oriented framing.


\subsection{Refining Specialist-Facing Interface prototype}

\begin{figure*}[t]
     \includegraphics[width=\textwidth]{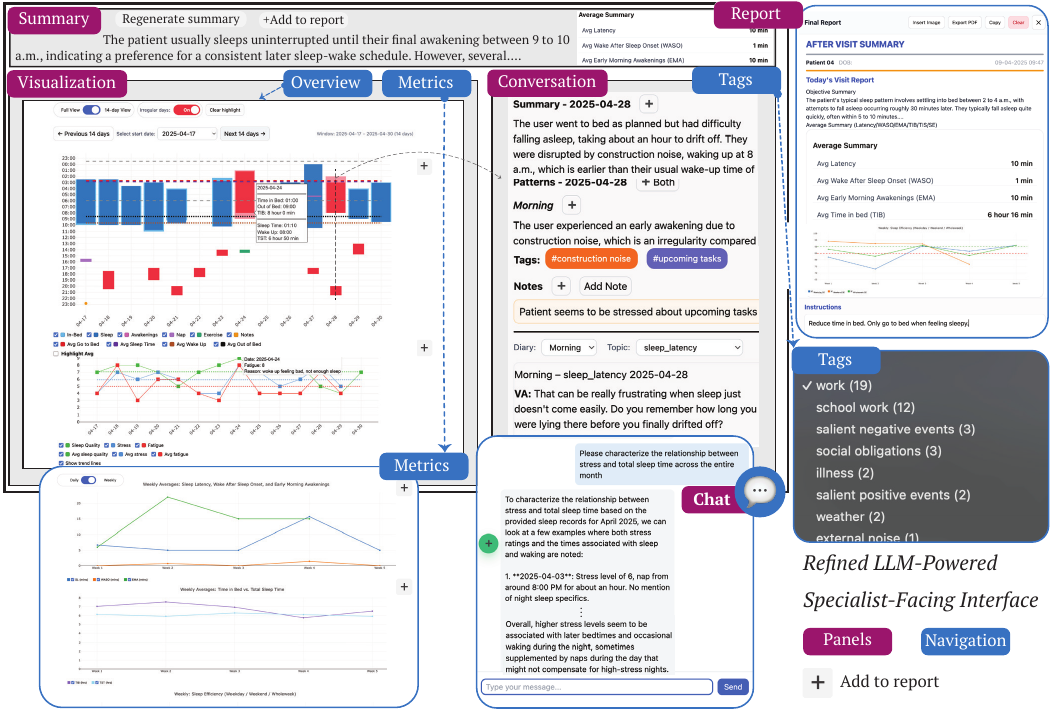}    
     \caption{Stage 2: Refined prototype for LLM-powered specialist-facing interface (part 2 and 3). The improvements are applied to the five panels: \textit{Summary}, \textit{Visualization}, \textit{Conversation}, \textit{Chat Bubble}, and \textit{Report}. We make the refined interface available for reviewers: \href{https://demo-sleepdairy-web.pages.dev/}{\underline{Specialist-facing analytics interface}}.}
     \Description{This figure shows the refined LLM-powered specialist-facing interface, which is divided into five parts to display the interface. There are also two types of titles: pink titles represent panels, and blue titles represent navigation. At the top left is a complete interface, which is divided into three parts: summary (pink title), visualization (pink title), and conversation (pink title). The summary (pink title) section of the interface has two buttons at the top, namely ``regenerate summary'' and ``+add to report''. below is a paragraph of text: the patient usually sleeps uninterrupted until their final awakening between 9 to 10 a.m., indicating a preference for a consistent later sleep-wake schedule. However, several... This section also includes a newly added table of average values on the right side.  
     On the bottom left of the summary (pink title) section is the visualization (pink title) section, which contains a bar chart, a line chart, and an overview (blue title) pointing to these two charts. On the right side of the visualization (pink title) section is the conversation (pink title) section. in the interface section, the text includes:  
     Summary – 2025-04-28  
     The user went to bed as planned but had difficulty falling asleep, taking about an hour to drift off. They were disrupted by construction noise, waking up at 8 a.m., which is earlier than their usual wake-up time of 10 a.m.  
     Patterns – 2025-04-28 Morning  
     The user experienced an early awakening due to construction noise, which is an irregularly compared event.  
     Tags: #construction noise #upcoming tasks  
     Notes: Patient seems to be stressed about upcoming tasks 
     Morning – sleep_latency 2025-04-28  
     VA: That can be really frustrating when sleep just doesn’t come easily. Do you remember how long you were lying there before you finally drifted off?  
     To the right of this part, there is a smaller section showing the report (pink title) section, and below it, another section with tags (blue title). This section shows a selection box containing tags: work (19), school work (12), salient negative events (3), social obligations (3), illness (2), salient positive events (2), weather (2), and external noise (1). The numbers in parentheses indicate the frequency of occurrence.  
     To the left of this, there are also two sections: metrics (blue title) and chat (pink title). metrics (blue title) displays two line charts. chat (pink title) displays a short conversation.  
     }
    \label{fig:stage2-refined-prototype} 
\end{figure*}

Based on the feedback from co-design workshops, we refined the prototype to incorporate the suggested features and improvements (Fig. \ref{fig:stage2-refined-prototype}). These refinements are reflected in the following parts on the interface (Fig. \ref{fig:stage2-refined-prototype}):

\begin{itemize} [leftmargin=*, nolistsep] 
    \item \textit{Summary Panel.}
Provides specialists with a quick overview of diary entries, including: (1) summary statistics (\eg averages and ranges across the diary period), and (2) LLM-generated key takeaways for the same period.
\item \textit{Visualization Panel.}
Includes interactive \textit{Overview} and \textit{Metrics} plots designed according to specialist feedback (Fig. \ref{fig:stage2-workshop-findings}). Specialists can add markers to annotate treatment plans or comments. Additionally, an LLM-powered agent can highlight irregular days on the overview plot upon  request.
\item \textit{Conversation Panel.} Provides access to diary entries by clicking data points in the \textit{Overview} plots. Entries can be filtered by topic (\eg sleep quality, fatigue, naps). Each entry includes an LLM-generated summary and highlights of patterns relative to a baseline (established using the first seven diary entries). The panel also displays \textit{Tags} generated by the LLM that capture internal and external disruptions or salient events; clicking a tag highlights all days with the same tag in \textit{Overview} graph. A dropdown menu above the panel provides an overview of all tags for the patient, along with their frequencies, to indicate recurring factors affecting sleep.

\item \textit{Chat Bubble.} Because specialists expressed satisfaction with this feature, it was retained was is in the refined design.

\item \textit{Report Panel.} Builds on the initial prototype by making it simple to add content from all the other panels. Specialists retain full control over report content, while commonly used information and all the plots are readily available to be included with a single click.

\end{itemize}

\begin{table}[!t]
\centering
\setlength{\tabcolsep}{3pt}
\renewcommand{\arraystretch}{0.95}
\caption{ \am{Results of Validation study: SUS score (out of 100) and feature ratings by participant (1–5; 1=Not at all useful, 2=Slightly useful, 3=Moderately useful, 4=Very useful, and 5=Extremely useful; included N/A - Did not use). Shading encodes usefulness ratings: red (< 2.5), orange (2.5–3.5, inclusive), and green (> 3.5) and SUS score: red (< 50), orange (60–70, inclusive representing ``OK'' usability), and green (> 70 representing ``Good'' usability). Mean (M) and standard deviations (SD) are provided as well.}}
\begin{tabular}{lrrrrrrrrrrr}
\toprule
Feature & C1 & P1 & F1 & F2 & F3 & \am{F4} & \am{F5} & \am{F6} & \am{F7} & M & SD \\
\midrule
Summary of sleep diaries & \heat{4} & \heat{5} & \heat{2} & \heat{4} & \heat{4} & \heat{5} & \heat{4} & \heat{3} & \heat{4} & \heat{3.89} & 0.93 \\
Overview plots & \heat{5} & \heat{3} & \heat{2} & \heat{5} & \heat{4} & \heat{5} & \heat{3} & \heat{4} & \heat{5} & \heat{4.00} & 1.12 \\
Treatment plan annotation on overview graph & \heat{4} & \heat{2} & \heat{3} & \heat{4} & \heat{4} & \heat{5} & \heat{5} & \heat{2} & \heat{4} & \heat{3.67} & 1.12 \\
Ability to add note to any day & \heat{3} & \heat{2} & \heat{3} & \heat{5} & \heat{4} & \heat{5} & N.A. & \heat{2} & \heat{5} & \heat{3.63} & 1.30 \\
Metric plots; averages, trendlines & \heat{5} & \heat{5} & \heat{3} & \heat{4} & \heat{4} & \heat{5} & \heat{3} & \heat{3} & \heat{5} & \heat{4.11} & 0.93 \\
Actual conversations; filtering & \heat{4} & \heat{2} & \heat{4} & \heat{5} & \heat{4} & \heat{2} & N.A. & \heat{2} & \heat{4} & \heat{3.38} & 1.19 \\
Tags for internal/external disruptions \& events & \heat{4} & \heat{3} & \heat{3} & \heat{5} & \heat{4} & \heat{5} & N.A. & \heat{4} & \heat{4} & \heat{4.00} & 0.76 \\
Highlight irregularities on overview graph & \heat{4} & \heat{3} & \heat{1} & \heat{5} & \heat{4} & \heat{5} & \heat{5} & \heat{3} & \heat{5} & \heat{3.89} & 1.36 \\
Chat with AI & \heat{5} & \heat{5} & \heat{2} & \heat{5} & \heat{4} & N.A. & N.A. & \heat{4} & \heat{3} & \heat{4.00} & 1.15 \\
Add to report & \heat{5} & \heat{5} & \heat{4} & \heat{5} & \heat{4} & \heat{5} & \heat{5} & \heat{4} & \heat{4} & \heat{4.56} & 0.53 \\
\hline
SUS score & \heatSUS{90.00} & \heatSUS{65.00} & \heatSUS{60.00} & \heatSUS{92.50} & \heatSUS{75.00} & \heatSUS{75.00} & \heatSUS{87.50} & \heatSUS{85.00} & \heatSUS{85.00} & \heatSUS{79.44} & 11.37 \\
\bottomrule
\end{tabular}

\label{tab:validation-study-results}
\end{table}

\begin{table*}[tb]
\centering
\caption{\am{Qualitative findings of validation study.}}
\label{tab:findings-evaluation}
\begingroup

\begin{tabular}{p{3.5cm} p{11.5cm}}

\textbf{Comments on system} & \textbf{Evidence from participant quotes} \\
\midrule
\midrule

1. Ease of patient data analysis &  F6: \tquotes{``I think the content of this tool is useful, I think it's a matter of refining certain features and improving the interface. I appreciate the ease it would provide to analyze patient data. ''}
\newline F7: \tquotes{``I think this would be a helpful tool for clinician.''}
\vspace{1mm}
\\
2. Useful features &
\textbf{Tags.} F6: \tquotes{``I found the \# [tag] component to be especially helpful with identifying trends. ''}
\newline \textbf{Add visuals to report.} P1: \tquotes{``I I like that you can add graphs and summaries to the final report. ''}
\newline \textbf{Bi-weekly summary.} P1: \tquotes{``I like that the average summary changes based on the 2 week period being looked at.''} 
\vspace{1mm}
\\
\hline
 \textbf{Feedback on system} & \textbf{Evidence from participant quotes} \\
\midrule
\midrule
 1. Inconsistencies in LLM-generated summaries & F1: \tquotes{``The inconsistencies with calculation made the summaries and graphs difficult to trust and interpret. Having accurate calculations of TIB, TST, WASO, and SE is so important for CBT-I. The other information is secondary.''} \newline F2:
 \tquotes{``Just make sure the AI summary is accurate; I’m not sure if it was every time''}
\vspace{1mm}
\\
2. Appropriateness of LLM-generated VA responses & F6: \tquotes{``[from conversation:]``I completely understand how that can be stressful. Deadlines can really weigh on you. Hopefully, you'll be able to wrap it up soon''. Some of these AI responses have unnecessary filler that could potentially elicit more stress or have the potential to provide misleading feedback.''}
\vspace{1mm}
\\
 3. Cluttered interface: information dense & P1: \tquotes{``The visual data for daytime variables (sleep quality, fatigue, stress) is a little busy and I think it makes the homepage too cluttered. This could be put on a third tab. The treatment notes are confusing to read---too cluttered.''} 
 \newline F6: \tquotes{``Would be nice to have a separate section dedicated for clinician notes rather than having to sift through to find them amongst other notes/data.''}
\vspace{1mm}
\\
\hline

 \textbf{Impact on BSM} & \textbf{Evidence from participant quotes} \\
 \midrule
\midrule
 1. Reducing patient burden &  F2: \tquotes{``I think it could be very useful for those on-the-go who are able to use it while getting ready in the morning.''} 
\vspace{1mm}
\\
 2. Capturing more subjective data & 
F1: \tquotes{``I like being able to quickly capture participants’ subjective report of their sleep. I think having more information would help clarify any unusual data points.''}
\vspace{1mm}
\\
 3. Reducing clinician burden & F6: \tquotes{``It could reduce the workload on clinicians that are
overwhelmed with demand currently, especially with pattern detection and generating clinical notes/summaries.'' }
\vspace{1mm}
\\
 4. Informing treatment planning & C1: \tquotes{``Increasing accuracy of sleep tracking and providing more contextual information to inform in-session discussion and treatment planning.'' }
\vspace{1mm}
\\
5. Patient-friendly reporting & F4: \tquotes{``Would provide more comprehensive sleep data and reports that are patient friendly. '' }
\newline F7: \tquotes{``Make it easy to provide visual graph to patient.''}
\vspace{1mm}
\\
 6. Improving patient adherence & F6: \tquotes{``While I don’t think it could fully automate the care, it could provide more daily support, reinforcement of education, and real-time data collection that could improve adherence. '' }
 \vspace{1mm}
\\
 7. Improve virtual/remote patient care & F5: \tquotes{``If a patient is using AI voice assistant, perhaps prompting of their sleep plan from sessions with behavioral sleep medicine specialists (\eg a reminder of the time they were supposed to get out of bed) could be helpful.''}
\newline  F6: \tquotes{``I think it would allow the opportunity to provide better care virtually, especially with structured therapies like CBTi.''}
\\
\hline
\end{tabular}
\endgroup
\end{table*}

\subsection{Validation Study with Sleep Specialists}
\label{sec:validation-study}
To validate the refined analytics tool, \am{we conducted an online study with five specialists (C1, P1, F1, F2, and F3 in Table~\ref{tab:participants}), all of whom participated in the broader design process, as well as with four participants who only interacted with the refined system (F4, F5, F6, F7) to gather additional feedback from naive users}. The goal was to assess the usability and perceived usefulness of the interface in clinical context. 

\subsubsection{Procedure}

Participants completed the study remotely via a Google Form that contained all study instructions. After providing informed consent, they first watched a short tutorial video of the interface. Each participant was then assigned a simulated patient case and asked to review the patient’s data using the tool as they would in a clinical setting. This process involved identifying key information, considering follow-up questions for the patient, and determining potential adjustments to the existing CBT-I treatment plan (a simulated plan was included in the overview graph). They also generated a patient report based on their review. 
Finally, they completed a post-study questionnaire.

\subsubsection{Metrics}

We evaluated the usability and usefulness of the interface as follows:

\begin{itemize} [leftmargin=*, nolistsep] 
    \item \textit{Usability score.} We calculated the System Usability Scale (SUS) score \cite{brooke1996sus}, a 10-item, five-point Likert scale. A score above 70 (range 0–100) is typically considered good (acceptable) usability \cite{bangor2009determining}.

    \item \textit{Usefulness of features.} Participants rated the usefulness of specific interface features in interpreting patient data and informing interventions in clinical settings (see Table \ref{tab:validation-study-results}) on a five-point Likert scale (1 = Not at all useful, 5 = Extremely useful; included N/A - Did not use).

    \item \textit{Open questions.} We asked participants two open-ended questions to gather: \am{1) general comments, suggestions, and feedback and 2) their perspectives on the impact of the conversational diary and analytics tool on BSM. Two coders first independently reviewed participant responses to the open text questions and than convened to extensively discuss and reach consensus on the: 1) comments on system, 2) feedback on system, and 3) impact of system on BSM.  }
    
\end{itemize}

\subsubsection{Results}
\label{sec:stage2-validation-results}

The SUS scores and feature usefulness ratings are presented in Table \ref{tab:validation-study-results}. Overall, participants rated the usability of system as good and found many of the  features useful. 
\am{However, F1’s and P1's ratings were lower than other participants.} \am{ For F1, primarily due to inconsistencies 
she observed between the overview summaries} (LLM-generated) and the statistics summary table (calculated from raw data) because the overview summary is generated by a language model and not computed. This mismatch led her to distrust other system outputs, which highlights the importance of accuracy in sleep metrics (Table \ref{tab:findings-evaluation}). 
\am{While P1 rated the tool lower because of cluttering (Table \ref{tab:findings-evaluation}).}

\am{Despite these concerns, participants recognized the system’s potential impact in BSM, detailed in Table \ref{tab:findings-evaluation}. They noted that the conversational diary tool and analytics tool could support multiple aspects of practice, including reducing patient and clinician burden and enhancing patient outcomes, such as adherence and remote care.}

\section{\am{Stage 3: Focus Group on Potential of CAs to Support Behavioral Sleep Medicine Broadly}}
\label{sec:co-design-workshops}
\am{After presenting the conversational diary and analytics tool and eliciting initial feedback, we engaged specialists in an in-depth discussion to further articulate their broader vision for the role of conversational agents in BSM.}


\subsection{Procedure and Data Analysis}
We conducted focus groups with the eight specialists (Table \ref{tab:participants}). Since all participants had already interacted with the conversational agent and \am{analytics tool} during the co-design workshops, we began directly with discussions on how they envisioned these tools supporting BSM. The focus groups were audio-recorded. \am{Each focus group followed an open discussion format and lasted between 30 and 60 minutes, 
with participant composition identical to that of the co-design workshops.}

\begin{figure*}[t]
     \includegraphics[width=\textwidth]{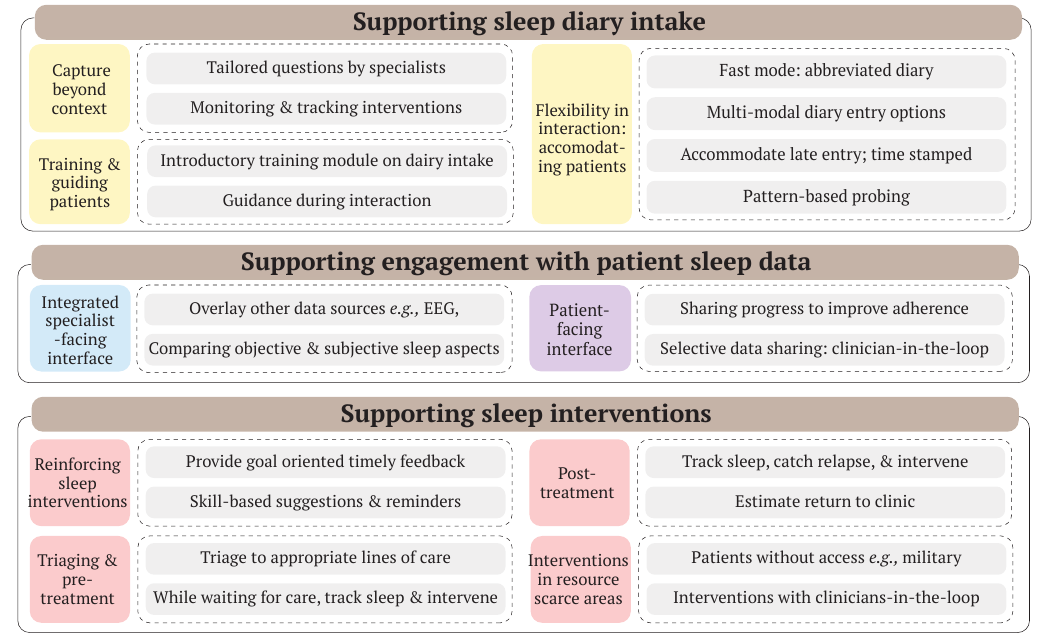}    
     \caption{Stage 3: Focus group findings. \am{Prospective look on how the conversational agents can support BSM in various aspects: (1) sleep diary intake, (2) engagement with patient sleep data, and (3) sleep interventions.} }
     \Description{This figure shows how the conversational agents can reshape BSM. The figure is divided into three sections from top to bottom: supporting sleep diary intake, supporting engagement with patient sleep data, and supporting sleep interventions.  
     The supporting sleep diary intake section has three parts:capture beyond context, training \& guiding patients, and flexibility in interaction: accommodating patients. Capture beyond context includes tailored questions by specialists and monitoring and tracking interventions. Training \& guiding patients includes introductory training module on diary intake and guidance during interaction. Flexibility in interaction: accommodating patients includes fast mode: abbreviated diary, multi-modal diary entry options, accommodate late entry; time stamped, and pattern-based probing.  
     The supporting engagement with patient sleep data section has two parts: integrated specialist-facing interface and patient-facing interface. Integrated specialist-facing interface includes overlay other data sources e.g., EEG and comparing objective \& subjective sleep aspects. Patient-facing interface includes sharing progress to improve adherence and selective data sharing: clinician-in-the-loop.  
     The supporting sleep interventions section has four parts: reinforcing sleep interventions, triaging \& pre-treatment, post-treatment, and interventions in resource scarce areas. Reinforcing sleep interventions includes provide goal oriented timely feedback and skill-based suggestions \& reminders. Triaging \& pre-treatment includes triage to appropriate lines of care and while waiting for care, track sleep \& intervene. Post-treatment includes track sleep, catch relapse, \& intervene and estimate return to clinic. Interventions in resource scarce areas includes patients without access e.g., military and interventions with clinicians-in-the-loop.  
}
    \label{fig:stage3-workshop-findings} 
\end{figure*}

We conducted \am{reflexive thematic analysis \cite{braun2023doing}} on the transcribed audio data from focus groups to identify how they envisioned the CAs being utilized in BSM. 
\am{Each transcript was analyzed independently by two members of the research team. First, they coded all transcripts and identified and grouped their thoughts into sub-themes and themes independently. The two authors then convened to extensively discuss and reach consensus on the the themes and sub-themes.}
We identified three main themes: (1) \am{supporting} sleep diary intake, (2) \am{supporting} engagement with patient data, and (3) \am{supporting} sleep interventions. Below, we present our findings (summarized in Fig. \ref{fig:stage3-workshop-findings}).

\subsection{Findings: \am{Supporting} Sleep Diary Intake} 
\label{sec:stage3-findings-sleep-diary-intake}

Our findings show that specialists envisioned conversational diaries as flexible tools that adapt to specialists' and  patients’ needs by moving beyond question-answer style interactions; we identified three sub-themes:

\subsubsection{Capture beyond context}

To enhance the patient understanding, participants envisioned the VA capturing more than just the context through simple follow ups. 

\begin{itemize} [leftmargin=*, nolistsep] 
\item \textit{Tailored questions by specialists.} 
Participants envisioned themselves being able to tailor questions for patients as needed, C3: \pquotes{``You could say, I want you to randomly ask their mood rating throughout the day, three times within these time periods, and a brief alarm would go off and they would just say, it's a 10.''}

\item \textit{Monitoring and tracking interventions.} 
Participants envisioned tracking patients’ use of treatment strategies, wanting to know whether techniques were applied in relevant situations. For example, P1 noted: \pquotes{``If they woke up during the night, I would want to know, what did you do during that time? Like, did you read, did you get out of bed, did you check your phone, anything like that,''} to better assess adherence and intervention effectiveness. 
\end{itemize}


Together, these ideas highlight specialists’ interest in conversational diaries that not only adapt questions to patient needs but also capture treatment adherence in real-world contexts.

\subsubsection{Flexibility in interaction: accommodating patients}
\label{sec:stage3-findings-sleep-diary-flexible-interaction}

To improve compliance, a conversational diary could lower barriers to completing diary by offering flexible modes of interaction. Participants envisioned the following possibilities: 

\begin{itemize} [leftmargin=*, nolistsep] 
    \item \textit{Fast mode: abbreviated diary.} A ``fast mode'' could allow patients to complete an abbreviated diary in 2–3 minutes when they are in a hurry, C2: \pquotes{``Where for an abbreviated diary, it can just go through in a way that might be more engaging, but succinct. And then an extended diary, the AI might probe a little more on observations that it's made.''}
    
    \item \textit{Multi-modal diary entry options.}
    For patients who cannot use their gadgets for example at work, multi-modal options may help ensure data capture. For example, patients could jot notes on paper during a break, take a picture, and later start a conversation with the VA to provide additional context, C2: \pquotes{``They literally write it on paper during lunch break and then they'll come home and type it in''} and \pquotes{``you don’t even have to enter [manually], you can take a picture.''}

    \item \textit{Accommodate late entry.} 
    If a patient misses a diary entry, the VA could check in later to allow completion---while still time stamping the entry so specialists can assess its validity, F1: \pquotes{``If there's like a day missing, maybe in the next time [the VA] might say, I noticed that you didn't answer the questions yesterday. Do you have a second to fill them?''}

    \item \textit{Pattern-based probing.}  
    The VA could use detected sleep patterns to flag irregularities and selectively probe for details---avoiding both overburdening patients with unnecessary questions and overlooking relevant ones in the interest of brevity.

\end{itemize}

Together, these ideas illustrate how a conversational VA could offer a more adaptive and patient-centered approach to diary intake, lowering compliance barriers while maintaining data quality and capturing rich contextual data.

\subsubsection{Training and guiding patients}
While a conversational diary allows flexible capture of context, patients may need support to use it appropriately. Participants emphasized the importance of both initial training and ongoing guidance:
\begin{itemize} [leftmargin=*, nolistsep] 
    \item \textit{Introductory training module on diary intake.}
    Patients may benefit from an on-boarding module delivered by the VA itself; helping them practice in completing diary entries and clarifying expectations, C3: \pquotes{``A chatbot training module, because these things that we're interested in are kind of confusing for people. So  helping them learn how to complete the diary correctly and what we want, that would be really cool.''} 

    \item \textit{Guidance during interaction.}
    Training alone may not be sufficient; patients may also need in-the-moment guidance to ensure accurate data capture. For instance, the VA could prompt for more precise numeric estimates when patients are vague, P2: \pquotes{``When my patient tells me, it's just a little bit, I still need [an estimate] 20 minutes, or is like 5 minutes?''} 

\end{itemize}

Together, these recommendations highlight the need for conversational VAs to train and guide patients for ensuring both accuracy and reducing barriers to adoption.


\subsection{Findings: \am{Supporting} Engagement with Patient Sleep Data}

Specialists have traditionally engaged with patient sleep data in fragmented and limited ways, often relying on manually processed diaries or reviewing information in isolation from other sources. When introduced to the potential of automated and AI-supported analytics, they envisioned more integrated systems that could support both specialists and patients. We identified two sub-themes:

\subsubsection{Integrated specialist-facing interface}
\label{sec:stage3-findings-visualization-intergrated}

Rather than treating diary entries as standalone artifacts, specialists envisioned interfaces that could layer multiple data streams. They emphasized two key aspects:

\begin{itemize} [leftmargin=*, nolistsep] 
    \item \textit{Overlaying other data sources.} The interface could integrate sleep diary data with other sources such as EEG, sleep trackers, and fitness trackers. It could also map health events by pulling from EMR records to automatically code significant changes, such as new medications or emergency visits, C3: \pquotes{``It could take data from the EMR [Electronic Medical Record] and that is tagged into some of these things potentially that would be useful, like they started a new medication, or prescribed a new medication that they got into an accident, went to the emergency room.''}
    \item \textit{Comparing objective and subjective sleep aspects.} Once subjective data (sleep diaries) and objective data (trackers) are combined, the system could help specialists analyze patterns across both to better understand patients' sleep. 
    
\end{itemize}

An integrated specialist-facing interface could provide a richer, more efficient view of patient sleep by combining subjective and objective data streams.

\subsubsection{Patient-facing interface}
\label{sec:stage3-findings-reporting}
Specialists envisioned to actively involve patients in reflecting on their progress more regularly. One way to achieve that would be through a patient-facing interface. They discussed two aspects:

\begin{itemize} [leftmargin=*, nolistsep] 
    \item \textit{Sharing progress to improve adherence.} 
    Reflecting on one’s own data can motivate patients and improve adherence, C1: \pquotes{``Probably across psychology, like eating for example, doing food diaries is an intervention in itself. I have definitely had some people who bring it back and say, wow, I gained so much insight, and I changed this and this, and I'm doing much better. When you can actually show them the data and say, on average look how much time in bed you're spending, this is inefficient, and we're just going to work on giving you better quality sleep. I think they buy into it a little bit more.''}

    \item \textit{Selective data sharing; specialist-in-the-loop.} On the flip side, too much detail may cause stress or anxiety. Patients can become hyperfocused on daily fluctuations and over-correct night to night. As C2 cautioned: \pquotes{``I think people can become hyperfocused on looking at night to night. But maybe having people be able to look at a weekly summary might be [better]. that way they're not like looking at it obsessively every day.''} 
    Thus, specialists may need to remain in the loop, deciding what patients can see and how frequently during the course of treatment.
    
\end{itemize}

These discussions underscore the double-edged nature of patient-facing interfaces: they can motivate adherence and reflection, but must be carefully managed to prevent over-engagement and anxiety.

\subsection{Findings: \am{Supporting} Sleep Interventions}
\label{sec:stage3-findings-interventions}
Conversational VAs have potential to go beyond supporting sleep diary intake and analytics; they could also play an active role before, during, and after CBT interventions. We identified four sub-themes:

\subsubsection{Reinforcing sleep interventions}

CAs could reinforce specialist-led interventions by detecting and tracking patients’ sleep patterns and delivering feedback grounded in the treatment plan. Participants highlighted two key opportunities:

\begin{itemize} [leftmargin=*, nolistsep] 
    \item \textit{Provide goal-oriented timely feedback.}
     VAs could deliver timely feedback based on diary conversations. This might include positive affirmations when patients meet their treatment goals, C3: \pquotes{``They show a pattern that we're asking for. If we said their new bedtime is 12 and they hit it 5 out of 7 days, it would be great for it to say, your doctor said, great job.''} They could also provide early warnings when negative patterns emerge, C1: \pquotes{``Catching it early and saying, Hey, I noticed that your middle of the night wake ups are getting longer. Is that something that we need to pay attention to?''}
    
    \item \textit{Skill-based suggestions and reminders.} 
    Beyond feedback, VAs could deliver reminders and suggestions aligned with the specialist’s treatment plan, P1: \pquotes{``It can pull some recommendations from the after visit summary, and there's a chance when they're doing the evening sleep diary to say, remember, don't go to bed until x time, as discussed with your therapist. Remember, get out of bed if you have been awake for longer than 10 minutes in the night.''} 
    Participants highlighted the importance of skill-based rather than outcome-based prompts, C1: \pquotes{``So based more on the skills again versus the outcome, you could say, Oh, you didn't sleep that well tonight, that's okay \dots Here's some things you can do if you feel like you want to take a nap \dots Go for a walk, call a friend, do something fun, and then tonight you'll sleep better, that kind of cheerleading.''}. Finally, participants also emphasized giving patients choice in the type and timing of reminders, as P2 proposed: \pquotes{``[VA can ask] Would you like a reminder of your weekly goal?''}

\end{itemize}




\subsubsection{Post-treatment}
CAs could also play a crucial role in supporting patients after treatment, helping to maintain gains and reduce the need for return visits---an important consideration given the months-long backlog for behavioral sleep care. Participants discussed several ways VAs could extend care post-treatment.

\begin{itemize} [leftmargin=*, nolistsep] 
    \item \textit{Track sleep, catch relapse, and intervene.} VA diary could be integrated into patients’ post-treatment routines in a lighter form (\eg abbreviated or less frequent entries) to provide ongoing monitoring. By tracking essential aspects of sleep, the VA could detect significant irregularities and flag early signs of relapse. When these patterns emerge, it could remind patients of the strategies they learned during treatment, C1: \pquotes{``It would be great if it kind of learned from the [patient] and over time helped, to catch lapses before it turned into a relapse. Like, oh, I noticed that the variability in your wake time is increasing, which we know increases risk of insomnia coming back. Here are the skills that you found helpful. Because that would probably reduce people returning to our clinic, which would help with wait times.''} 

    \item \textit{Estimate return to clinic.}
    By monitoring post-treatment patterns, CAs could help identify when professional care is needed, flagging a return to clinic if specialist-recommended interventions prove ineffective.

\end{itemize}

\subsubsection{Triaging and pre-treatment}

Similar to post-treatment care, participants also highlighted the potential role of CAs in the pre-treatment phase. They discussed two ways it could help streamline care and improve efficiency:
\begin{itemize} [leftmargin=*, nolistsep] 
    
    \item \textit{Triage to appropriate lines of care.}
    Participants noted that patients are often bounced between different care providers before reaching the appropriate treatment, C1: \pquotes{``I think because people get bounced around a lot, sometimes they come to us and we're like, actually, you need to go back to sleep medicine because it's an insomnia problem\dots If everybody who expressed interest or got scheduled did some monitoring, we could figure out, okay, this person is going to do [better] more likely in this pathway, and divert them there. I think that would be a way more efficient use of resources.''} CAs could support this process by beginning to phenotype patients---tracking sleep patterns through diaries and integrating data from EEG or other trackers---so they can be triaged into the most appropriate line of care, C1: \pquotes{``We could eventually start phenotyping people and, like, triaging them into the appropriate lines of care.''}

    \item \textit{While waiting for care, track sleep and intervene.}    
    Participants envisioned VAs providing interim support during long wait times to care. They could begin tracking their sleep with VA diaries and receive simple, skill-based interventions for less complex insomnia cases, C1: \pquotes{``I think eventually, there's low hanging fruit people, where just a couple interventions definitely help, where if it learns enough to give them some specific recommendations, and they're highly motivated and it's a simple case, then great, they're good to go. They don't have to come in.''} 
    In addition, all data collected during the waiting period would provide specialists with a rich baseline, improving the efficiency and quality of initial sessions.  

\end{itemize}

Together, these ideas highlight how conversational VAs could improve efficiency in the pre-treatment phase by triaging patients to the right pathway, offering interim support, and establishing a baseline of sleep data to accelerate treatment once care begins.



\subsubsection{Intervention in resource scarce areas}

Participants also discussed the broader potential of CAs to deliver behavioral sleep interventions in areas with limited access. They highlighted both the opportunities and the safeguards needed to make such use cases feasible:

\begin{itemize} [leftmargin=*, nolistsep] 
    \item \textit{Patients without access. } Poor sleep health and insomnia are especially prevalent in populations such as armed forces deployed in remote locations, where access to specialists is limited Participants envisioned CAs providing structured interventions in these settings.  

    \item \textit{Interventions with specialists-in-the-loop.}

    While CAs can deliver components of behavioral sleep interventions, participants emphasized the need for specialist involvement in design and oversight to keep interventions safe and aligned with clinical practices.
 
\end{itemize}


Stage 3 highlighted specialists’ vision for how CAs could further support BSM. They imagined flexible and adaptive sleep diary intake, integrated ways of engaging with patient data, and new opportunities to support interventions before, during, and after treatment, including in resource-scarce areas.

\section{General Discussion and Conclusion}

\am{In this work, we explored how conversational agents might complement existing sleep diary practices in Behavioral Sleep Medicine in collaboration with specialists through design of sleep diary intake and specialist-facing analytics.} We co-designed and developed an LLM-powered VA sleep diary, along with a specialist-facing analytics tool, with the goal of reducing burden on both patients and specialists and solving issues limited context, \am{identified through interviews with specialists}. Through this design process, specialists began to envision how conversational agents could extend beyond diary intake \am{to support broader practices in BSM}.





\subsection{\am{Conversational Agents as a New Complementary Modality for Self-Reporting: Design Implications}}





\am{The introduction of conversational agents---specifically voice assistants powered by LLMs---represents a shift in the interaction paradigm for sleep diary intake and data analytics. Conversational agents offer an alternative method of diary collection that may have potential to reduce patient burden, support richer contextual capture, make specialist analysis more efficient and complement existing clinical workflows, as expressed by specialists.} 

\subsubsection{Conversational dairy: Towards flexible, adaptive self-reporting and monitoring}
Specialists in our study recognized that a conversational diary could move beyond the rigid structure of traditional formats by adaptively probing for context. They envisioned it not only as a tool for tracking sleep patterns but also for monitoring patients’ use of treatment strategies and their effectiveness (Section \ref{sec:stage3-findings-sleep-diary-intake}). 
\am{Such rich, experience-based data could inform ongoing interventions, highlighting the dual role of self-report tools in capturing both behaviors and adherence---going beyond what traditional diary modalities typically capture in routine practice \cite{carney2012consensus}.
}

Specialists further highlighted how the flexible interaction style of conversational diaries (Section \ref{sec:stage3-findings-sleep-diary-flexible-interaction}) could reduce patient burden and increase compliance. For instance, features such as embedded training, guidance modules, and fast-mode were seen as particularly helpful for lowering barriers to adoption and completion. Specialists also noted that location is often a bottleneck for diary completion---for instance, at work or while traveling. In our deployment, we observed that students used the Alexa app on their phones while away from home, extending use beyond the at-home smart speaker.
This cross-platform nature of conversational VAs such as Alexa, Siri, and Google Assistant—now available on smartphones, watches, glasses, and even cars—points toward a new medium of situated diary intake. Portability creates both opportunities and challenges: conversational diaries can integrate into daily routines more seamlessly than paper or static apps, but designers must ensure consistency across contexts and address privacy concerns in shared environments. Multi-modal input can further support this flexibility, as modern CAs offer both voice and text entry. 

\begin{adjustwidth}{0.3cm}{0pt}

\noindent
\am{\textbf{Design Implication 1: }
\textit{Enable seamless cross-platform, context-sensitive diary interactions that adapt to users’ environments while maintaining consistency and privacy.}}

\noindent
\am{\textbf{Design Implication 2:}
\textit{Create interaction flows that adjust diary intake depth, timing, and modality to reduce burden and sustain user engagement.}}

\end{adjustwidth}


\subsubsection{\am{Specialist-facing analytics tools:} Towards dynamic, conversational engagement with patient data}
To address specialist burden---identified in interviews (Section \ref{sec:interviews}) and supported by prior work\cite{islind2019shift, schmitz2022towards}---our participants went beyond improving efficiency through an automated analytics tool: they saw potential of conversational interfaces for patient data to enhance engagement with patient data and unlock new interaction possibilities. Enabled by an LLM-powered agent, conversational querying shifts the workflow from passively reviewing static summaries and plots to actively exploring large volumes of conversational records in an open-ended, flexible manner. In our work, we used zero-shot prompting with GPT-4o to demonstrate the feasibility and potential benefits of this paradigm.
Recent research is moving toward making conversational querying more robust by integrating diverse data sources. For instance, the openCHA framework incorporates multi-modal patient data, knowledge bases, and analysis models to enable personalized querying, generate patient reports, and support follow-up questions \cite{abbasian2025conversational}. 
Our participants, however, emphasized that AI agents should go beyond querying alone and be more tightly integrated into visualizations (\eg highlighting irregularities) while still supporting conversational exploration of patterns, trends, and specific questions (Fig. \ref{fig:stage2-workshop-findings}). They also expressed interest in more robust statistical analyses, such as correlations between sleep and stress, which frameworks such as openCHA could support through dedicated analytical modules \cite{abbasian2025conversational}. 
Building on this, recent frameworks \cite{abbasian2025conversational,merrill2024transforming} demonstrate how combining data from diverse sources can provide personalized insights. For instance, PHIA \cite{merrill2024transforming} shows how personal tracker data can be leveraged to give users fluid, conversational access to health insights.
\am{In line with this vision, participants also envisioned integrating subjective diary data with objective sources such as actigraphs, smartwatches, or EEG recordings (Section \ref{sec:stage3-findings-visualization-intergrated}), suggesting that they viewed CAs as a complementary tool to support their existing practice.} Such integration could strengthen clinical practice while advancing the field’s understanding of sleep, as subjective and objective data provide complementary perspectives \cite{silva2007relationship, schmitz2022towards}. At the same time, caution is warranted with wearable technologies such as Fitbit, as their clinical validity has not been firmly established \cite{jang2023chatbotfitbit, silva2007relationship}. Thus, analytics tools are not just about efficiency, but about enabling richer engagement with patient data to enhance BSM practice and knowledge.

\begin{adjustwidth}{0.3cm}{0pt}

\noindent
\am{\textbf{Design Implication 3: }
\textit{Embed AI-assisted analysis in visualizations to support a dialogic, back-and-forth interpretation, allowing specialists to interactively identify patterns and irregularities.}}

\noindent
\am{\textbf{Design Implication 4:}
\textit{Complement existing objective data analytics (e.g., actigraphs and sensors) with subjective diary data to support deeper clinical sensemaking.}}

\noindent
\am{\textbf{Design Implication 5:}
\textit{Clearly communicate the reliability and constraints of subjective and objective data sources to support informed clinical interpretation.}}

\end{adjustwidth}


\subsubsection{\am{Analytics tools for patients: Towards safe, continuous patient engagement and education}}

While our work supported prior work showing the benefits of patient-facing interfaces such as adherence and motivation \cite{almahmud2022sleepapps, philip2020covidagent, kuhn2016cbticoach, espie2012sleepio}, we uncovered the flip side:
specialists in our study emphasized the risks of presenting too much information in unmediated ways (Sections \ref{sec:stage2-workshop-findings-reporting} and \ref{sec:stage3-findings-reporting}). Excessive access to daily data, they noted, can cause anxiety and lead patients to over-correct or under-correct their behaviors which can undermine the treatment. 
Specialists also noted that patient preferences vary: some want detailed graphs and metrics, while others prefer summaries. This underscores the need to involve not only patients (the target users) but also behavioral specialists in designing interfaces for patients. Tailoring information to a patient’s anxiety, goals, and treatment progress---under the guidance of specialists--- can make these tools more effective while keeping them safe and clinically valid.

\begin{adjustwidth}{0.3cm}{0pt}

\noindent
\am{\textbf{Design Implication 6:}
\textit{Support adjustable views, from high-level summaries to more detailed metrics, to accommodate different patient comfort levels and information-processing styles to avoid anxiety.}}

\noindent
\am{\textbf{Design Implication 7:}
\textit{Design specialist-in-the-loop interface for patients to ensure that they remain supportive, safe, and aligned with therapeutic goals rather than overwhelming or counterproductive.}}

\end{adjustwidth}



\subsubsection{Integration of LLMs: Towards safe, reliable healthcare solutions.}
While specialists recognized the promise of LLM-powered conversational agents, they also raised concerns about inconsistencies and errors in generative models (Section \ref{sec:stage2-validation-results}), known issues with LLMs along with hallucinations, and lack of transparent reasoning \cite{koubaa2023exploring}. In health-critical settings, such issues can carry serious risks \cite{ong2024ethical, hastings2024preventing}. 
\am{Our validation study surfaced these concerns directly: discrepancies between calculated sleep metrics and LLM-generated summaries led participants to question reliability, underscoring the need to catch, correct, and transparently communicate such errors. Although our preliminary analysis of 20\% of the conversational data using human-coded validation showed that core sleep parameters can be reliably extracted by LLMs, validation of LLM-generated summaries and the CA’s conversational responses still needs to be systematically established.}

\am{To move toward safe and trustworthy systems, future work must incorporate architectures that improve consistency for essential sleep metrics, such as tool-calling approaches \cite{paranjape2023art}, which rely on deterministic metric calculations rather than free-form generation and therefore offer a promising pathway for reducing hallucinations. Moreover, concrete mitigation strategies should be implemented, including prompt guardrails that constrain model behavior; evidence-linked responses that point to specific diary segments or data sources; model rationales that explain how outputs were generated; and verification workflows (e.g., flagging output for review \cite{ahmad2023creating}) that allow clinicians to review, correct, or override system outputs (human-in-the-loop \cite{akheel2025guardrails}).}
Further research is needed on benchmarking frameworks, risk-assessment methodologies, enriched training datasets, and robust safeguards to ensure safe deployment of LLM-powered VAs in healthcare \cite{ong2024ethical, hastings2024preventing}. Lastly, we should explore strategies for communicating clearly with both specialists and patients about which outputs are AI-generated and how they should be interpreted. To ensure these solutions are safe and reliable, stakeholders, especially healthcare professionals, must be involved throughout the design process.

\begin{adjustwidth}{0.3cm}{0pt}

\noindent
\am{\textbf{Design Implication 8:}
\textit{Implement tool-calling and deterministic computation for clinically relevant metrics to ensure accuracy and consistency.}}

\noindent
\am{\textbf{Design Implication 9:}
\textit{Integrate human-in-the-loop verification pathways that allow clinicians to easily review, edit, and override model outputs, especially in cases where discrepancies or uncertainties are detected.}}

\noindent
\am{\textbf{Design Implication 10:}
\textit{Provide specialists and patients with clear indicators of which outputs are AI-generated and link each summary or recommendation to the specific diary segments or data sources that support it.}}

\end{adjustwidth}

\subsection{Extending Care Beyond The Clinic: CAs to Support BSM at-Home}

Personalized at-home care is becoming an inevitable reality in BSM due to: 1) potential ability of technology to deliver CBT interventions at-home \cite{chen2024dozzz, groninger2025voicecancer, almzayyen2022voiceagents, almahmud2022sleepapps, philip2020covidagent, kuhn2016cbticoach, espie2012sleepio} and 2) a need due to overburdened BSM provider systems\cite{mustafa2005sleep, taylor2017internet, germain2021enhancing}.  
First, specialists saw opportunities and explored feasibility for CAs to be applied across the continuum of care---pre-, during, and post-treatment---to extend BSM beyond the clinic. 
For instance, CAs can triage patients into appropriate lines of care, support interventions by providing timely feedback and reminders, and prevent return to clinic by helping patients track their sleep post-treatment (Section \ref{sec:stage3-findings-interventions}). 

Second, our participants noted the potential of CAs to support interim care \cite{shah2025effects} during long wait times (C1: \pquotes{``Our wait list right now, people are booking it to, I think, March of next year [7 months].''}).
They pointed out that CAs could be particularly valuable in resource-limited contexts or among populations with limited access to care. For instance, for military personnel and veterans where insomnia is highly prevalent and resources are scarce \cite{mustafa2005sleep, taylor2017internet, germain2021enhancing}. Additionally, CAs can support overburdened healthcare systems since sleep specialists remain a scarce resource (C1: \pquotes{``Not many people do sleep behavioral work, unfortunately, so it's hard to find somebody.'')}, often concentrated in urban academic medical centers \cite{sivertsen2013future, thomas2016behavioral, perlis2008can, germain2021enhancing}.  
This vision underscores the broader promise of CAs: scalable, accessible, and cost-effective support for sleep health, similar to other digital interventions \cite{drake2016promise, sanabria2025active}. Beyond efficiency, these tools may also promote healthcare equity, as AI-driven sleep technologies can expand access and help reduce disparities in care among underserved populations \cite{bandyopadhyay2024strengths}.

For efficacy, however, specialists emphasized the importance of keeping providers in the loop, similar to prior work \cite{islind2019shift, cerna2020changing, tiase2020integration, sadhu2023dashboard}. Our work provides further insights into how to achieve that. For instance, specialists envisioned a division of responsibilities; CAs support continuous tasks---tracking patterns, reinforcing behaviors, and delivering reminders---while specialists step in as needed to monitor progress and adjust interventions. This balance ensures that CAs act as supportive extensions of clinical care rather than independent agents.

\subsection{Broader Implications for Behavioral Medicine and Healthcare}

Beyond sleep medicine, self-reporting remains the most common strategy for tracking patient behaviors during treatment and informing behavioral interventions \cite{arigo2019history}. Diaries are widely used across behavioral medicine—for example, in tracking mood \cite{aan2012mood}, pain \cite{stone2003patient}, eating behaviors \cite{burke2011self}, and substance use \cite{shiffman2009ecological}---yet they face persistent challenges, including low compliance, recall bias, backfilling, and limited contextual capture.
On the other hand, in chronic disease management, self-reporting supports self-management of conditions such as hypertension and diabetes \cite{hunt2015technology} and has been shown to improve treatment compliance \cite{van1999patient}. While involving care teams in self-reporting improves disease management \cite{tang2013online, hunt2015technology}, day-to-day monitoring tools are rarely integrated into practice due to the burden they place on physicians and patients. As a result, patients often rely on memory to report key behaviors (\eg diet, exercise), which is prone to error.
Our findings highlight the potential of proactive, adaptive conversational diaries to reduce burden for both patients and providers, improve compliance, and mitigate recall-related errors by capturing information closer to the moment of experience. Moreover, combining subjective reports (\eg food diaries) with objective physiological data (\eg glucose levels) could provide a more complete picture of health \cite{nes2013web, aarsand2010mobile}. Such integration---enabled for example through specialist-facing interfaces (Section \ref{sec:stage3-findings-visualization-intergrated})---could strengthen decision-making and support personalized treatment. Extending conversational diary and \am{analytics tools} into these domains has the potential to \am{support} existing practices and enable more tailored, behavior-focused care.




\section{Limitations and Future Work}
While our work provides empirical insights into how conversational agents may support behavioral sleep medicine, a large-scale study in real clinical settings is needed to validate the efficacy of the conversational diary and analytics tool. Our data collection was conducted with university students who may experience sleep difficulties but were not diagnosed patients; future work should test the tool in clinical populations and across more diverse groups. \am{For instance, what a VA may or may not say during intake can vary in important ways and may have significant impact on real patients, as pointed out by F6 in Table \ref{tab:findings-evaluation} (\textit{``appropriateness of LLM-generated responses''}). }
\am{Although we engaged clinicians from three healthcare systems and at different career stages to capture varied perspectives, our sample size is small and further studies with larger and more diverse clinician samples are needed---particularly to evaluate the applicability of the analytics tool.}

Furthermore, our work did not examine the potential risks of applying LLMs in sleep data interpretation, such as hallucinations, bias, and reproducibility concerns. Future research should investigate these risks and develop strategies to mitigate them, ensuring interactions remain safe and reliable for clinical use. Moreover, the use of LLMs in this context raises additional challenges around data protection and broader ethical issues related to storing and processing personal health-related conversations. Future work is needed to ensure that conversational agents in behavioral medicine, and healthcare more broadly, are developed with strong safeguards to protect patient safety and privacy.

\section*{Declaration of generative AI and AI-assisted technologies in the writing process}

During the preparation of this work the authors used ChatGPT in order to cut down repetitions and improve readability and language. After using this tool/service, the authors reviewed and edited the content as needed and take full responsibility for the content of the publication.

\newpage

\bibliographystyle{ACM-Reference-Format}
\bibliography{references}

@article{mahmood2025user,
  title={User interaction patterns and breakdowns in conversing with llm-powered voice assistants},
  author={Mahmood, Amama and Wang, Junxiang and Yao, Bingsheng and Wang, Dakuo and Huang, Chien-Ming},
  journal={International Journal of Human-Computer Studies},
  volume={195},
  pages={103406},
  year={2025},
  publisher={Elsevier}
}

@article{stone2002capturing,
  title={Capturing momentary, self-report data: A proposal for reporting guidelines},
  author={Stone, Arthur A and Shiffman, Saul},
  journal={Annals of Behavioral Medicine},
  volume={24},
  number={3},
  pages={236--243},
  year={2002},
  publisher={Springer}
}

@article{harris2009redcap,
  title={Research electronic data capture (REDCap)—A metadata-driven methodology and workflow process for providing translational research informatics support},
  author={Harris, Paul A and Taylor, Robert and Thielke, Robert and Payne, Jonathon and Gonzalez, Nathaniel and Conde, Jose G},
  journal={Journal of Biomedical Informatics},
  volume={42},
  number={2},
  pages={377--381},
  year={2009},
  publisher={Elsevier}
}

@article{bolger2003diary,
  title={Diary methods: Capturing life as it is lived},
  author={Bolger, Niall and Davis, Angelika and Rafaeli, Eshkol},
  journal={Annual Review of Psychology},
  volume={54},
  number={1},
  pages={579--616},
  year={2003},
  publisher={Annual Reviews}
}

@article{smyth2003ema,
  title={Ecological momentary assessment research in behavioral medicine},
  author={Smyth, Joshua M and Stone, Arthur A},
  journal={Journal of Happiness Studies},
  volume={4},
  number={1},
  pages={35--52},
  year={2003},
  publisher={Springer}
}

@article{shiffman2008ema,
  title={Ecological momentary assessment},
  author={Shiffman, Saul and Stone, Arthur A and Hufford, Michael R},
  journal={Annual Review of Clinical Psychology},
  volume={4},
  pages={1--32},
  year={2008},
  publisher={Annual Reviews},
  doi={10.1146/annurev.clinpsy.3.022806.091415}
}

@article{mohr2017behavioral,
  title={Behavioral intervention technologies: Evidence review and recommendations for future research in mental health},
  author={Mohr, David C and Riper, Heleen and Schueller, Stephen M},
  journal={General Hospital Psychiatry},
  volume={49},
  pages={9--20},
  year={2017},
  publisher={Elsevier}
}

@article{mohr2014bit,
  author    = {Mohr, David C. and Burns, Michelle N. and Schueller, Stephen M. and Clarke, Gregory and Klinkman, Michael},
  title     = {The Behavioral Intervention Technology Model: An Integrated Conceptual and Technological Framework for eHealth and mHealth Interventions},
  journal   = {Journal of Medical Internet Research},
  year      = {2014},
  volume    = {16},
  number    = {6},
  pages     = {e146},
  doi       = {10.2196/jmir.3077},
  url       = {https://www.jmir.org/2014/6/e146/}
}

@article{riley2011models,
  author    = {Riley, William T. and Rivera, Daniel E. and Atienza, Audie A. and Nilsen, Wendy and Allison, Susannah M. and Mermelstein, Robin},
  title     = {Health behavior models in the age of mobile interventions: Are our theories up to the task?},
  journal   = {Translational Behavioral Medicine},
  year      = {2011},
  volume    = {1},
  number    = {1},
  pages     = {53--71},
  doi       = {10.1007/s13142-011-0021-7},
  url       = {https://pmc.ncbi.nlm.nih.gov/articles/PMC3142960/}
}

@article{arigo2025behavioral,
  author    = {Arigo, Danielle and Hekler, Eric and Pagoto, Sherry and Spring, Bonnie and King, Abby and Nahum-Shani, Inbal},
  title     = {The recent history and near future of digital health in behavioral medicine},
  journal   = {Journal of Behavioral Medicine},
  year      = {2025},
  volume    = {48},
  number    = {1},
  pages     = {1--18},
  doi       = {10.1007/s10865-024-00526-x},
  url       = {https://link.springer.com/article/10.1007/s10865-024-00526-x}
}

@article{ghadi2025wearable,
  author    = {Ghadi, Yasin and Rahman, S. and Alshamrani, S. and Albarrak, A.},
  title     = {Integration of wearable technology and artificial intelligence in digital health for remote patient care},
  journal   = {Journal of Cloud Computing},
  year      = {2025},
  volume    = {14},
  number    = {1},
  pages     = {1--20},
  doi       = {10.1186/s13677-025-00759-4},
  url       = {https://journalofcloudcomputing.springeropen.com/articles/10.1186/s13677-025-00759-4}
}

@article{canali2022wearables,
  author    = {Canali, Chiara and Bruzzone, Andrea and Gamberini, Lisa and Rizzo, Alessandro and Tzafestas, Elias},
  title     = {Challenges and recommendations for wearable devices in digital health: Data quality, interoperability, health equity, and fairness},
  journal   = {Digital Health},
  year      = {2022},
  volume    = {8},
  pages     = {1--15},
  doi       = {10.1177/20552076221105125},
  url       = {https://www.researchgate.net/publication/363563069}
}

@article{walker2021sleep,
  author    = {Walker, Matthew P.},
  title     = {Sleep Essentialism},
  journal   = {Brain},
  year      = {2021},
  volume    = {144},
  number    = {3},
  pages     = {697--699},
  doi       = {10.1093/brain/awab044},
  url       = {https://doi.org/10.1093/brain/awab044},
  publisher = {Oxford University Press}
}

@article{benjafield2019osa,
  author    = {Benjafield, Adam V. and Ayas, Najib T. and Eastwood, Peter R. and Heinzer, Raphael and Ip, Mary S.M. and Morrell, Mary J. and Nunez, Carlos M. and Patel, Sanjay R. and Penzel, Thomas and Pépin, Jean-Louis and Malhotra, Atul},
  title     = {Estimation of the global prevalence and burden of obstructive sleep apnoea: a literature-based analysis},
  journal   = {The Lancet Respiratory Medicine},
  year      = {2019},
  volume    = {7},
  number    = {8},
  pages     = {687--698},
  doi       = {10.1016/S2213-2600(19)30198-5},
  url       = {https://doi.org/10.1016/S2213-2600(19)30198-5}
}

@inproceedings{vacaretu2019digitalsleep,
  author    = {Văcăreţu, Nicolae and Damsa, Cosmin and Vancea, Andreea and Pădure, Mihai},
  title     = {Towards a Digital Sleep Diary Standard},
  booktitle = {2019 E-Health and Bioengineering Conference (EHB)},
  year      = {2019},
  pages     = {1--4},
  doi       = {10.1109/EHB47216.2019.8969887},
  url       = {https://ieeexplore.ieee.org/document/8969887}
}

@article{blake2010online,
  author    = {Blake, Jacqueline and Kerr, Donald V.},
  title     = {Development of an Online Sleep Diary for Physician and Patient Use},
  journal   = {Knowledge Management \& E-Learning},
  year      = {2010},
  volume    = {2},
  number    = {2},
  pages     = {188--202}
}

@article{carney2012consensus,
  author    = {Carney, Colleen E and Buysse, Daniel J and Ancoli-Israel, Sonia and Edinger, Jack D and Krystal, Andrew D and Lichstein, Kenneth L and Morin, Charles M},
  title     = {The Consensus Sleep Diary: Standardizing Prospective Sleep Self-Monitoring},
  journal   = {Sleep},
  year      = {2012},
  month     = feb,
  volume    = {35},
  number    = {2},
  pages     = {287--302},
  doi       = {10.5665/sleep.1642},
  pmid      = {22294820},
  pmcid     = {PMC3250369}
}

@article{almahmud2022sleepapps,
  author    = {Al Mahmud, Abdullah and Wu, Jun and Mubin, Omar},
  title     = {A Scoping Review of Mobile Apps for Sleep Management: User Needs and Design Considerations},
  journal   = {Frontiers in Psychiatry},
  year      = {2022},
  month     = oct,
  day       = {18},
  volume    = {13},
  eid       = {1037927},
  doi       = {10.3389/fpsyt.2022.1037927},
  pmid      = {36329917},
  pmcid     = {PMC9624283},
  publisher = {Frontiers}
}

@article{philip2020covidagent,
  author    = {Philip, Pierre and Dupuy, L and Morin, Charles M and de Sevin, E and Bioulac, S and Taillard, J and Serre, F and Auriacombe, M and Micoulaud-Franchi, Jean-Arthur},
  title     = {Smartphone-Based Virtual Agents to Help Individuals With Sleep Concerns During COVID-19 Confinement: Feasibility Study},
  journal   = {Journal of Medical Internet Research},
  year      = {2020},
  month     = dec,
  day       = {18},
  volume    = {22},
  number    = {12},
  eid       = {e24268},
  doi       = {10.2196/24268},
  pmid      = {33264099},
  pmcid     = {PMC7752183},
  publisher = {JMIR Publications}
}

@article{jang2023chatbotfitbit,
  author    = {Jang, H and Lee, S and Son, Y and Seo, S and Baek, Y and Mun, S and Kim, H and Kim, I and Kim, J},
  title     = {Exploring Variations in Sleep Perception: Comparative Study of Chatbot Sleep Logs and Fitbit Sleep Data},
  journal   = {JMIR mHealth and uHealth},
  year      = {2023},
  month     = nov,
  day       = {21},
  volume    = {11},
  eid       = {e49144},
  doi       = {10.2196/49144},
  pmid      = {37988148},
  pmcid     = {PMC10698662},
  publisher = {JMIR Publications}
}

@inproceedings{chen2024dozzz,
  author    = {Chen, Shanshan and Markopoulos, Panos and Hu, Jun},
  title     = {Dozzz: Exploring the Feasibility of a Voice-Based Sleep Diary for Children},
  booktitle = {Proceedings of the 37th International BCS Human-Computer Interaction Conference},
  year      = {2024},
  publisher = {BCS Learning \& Development Ltd},
  address   = {Swindon, GBR},
  pages     = {100--114},
  numpages  = {15},
  doi       = {10.14236/ewic/BCSHCI2024.10},
  url       = {https://doi.org/10.14236/ewic/BCSHCI2024.10},
  location  = {University of Central Lancashire (UCLan)},
  series    = {BCS HCI '24}
}

@inproceedings{chen2025multimodal,
  author       = {Chen, Shanshan and Hu, Jun and van Iterson, Hannah Christina and Fang, Ning and Markopoulos, Panos},
  title        = {"Did you sleep well?": A Multimodal Sleep Diary for Sustained Self-Reporting by Children},
  booktitle    = {Proceedings of the 2025 CHI Conference on Human Factors in Computing Systems (CHI '25)},
  year         = {2025},
  articleno    = {1178},
  numpages     = {17},
  publisher    = {Association for Computing}
}

@article{groninger2025voicecancer,
  author    = {Groninger, H and Arem, H and Ayangma, L and Gong, L and Zhou, E and Greenberg, D},
  title     = {Development of a Voice-Activated Virtual Assistant to Improve Insomnia Among Young Adult Cancer Survivors: Mixed Methods Feasibility and Acceptability Study},
  journal   = {JMIR Formative Research},
  year      = {2025},
  volume    = {9},
  eid       = {e64869},
  doi       = {10.2196/64869},
  pmid      = {40063947},
  pmcid     = {PMC11933750},
  publisher = {JMIR Publications}
}

@article{almzayyen2022voiceagents,
  author    = {Almzayyen, Abdalsalam and Evia, Angel and Coronato, Nick and Boukhechba, Mehdi},
  title     = {Voice-Based Conversational Agents for Self-Reporting Fluid Consumption and Sleep Quality},
  journal   = {arXiv preprint arXiv:2202.02186},
  year      = {2022},
  month     = feb,
  day       = {4},
  doi       = {10.48550/arXiv.2202.02186},
  url       = {https://arxiv.org/abs/2202.02186}
}

@article{kuhn2016cbticoach,
  author    = {Kuhn, Eric and Weiss, Brandon J and Taylor, Kelly L and Hoffman, Justin E and Ramsey, Katherine M and Manber, Rachel and Gehrman, Philip and Crowley, Jennifer J and Ruzek, Josef I and Trockel, Mickey},
  title     = {CBT-I Coach: A Description and Clinician Perceptions of a Mobile App for Cognitive Behavioral Therapy for Insomnia},
  journal   = {Journal of Clinical Sleep Medicine},
  year      = {2016},
  month     = apr,
  day       = {15},
  volume    = {12},
  number    = {4},
  pages     = {597--606},
  doi       = {10.5664/jcsm.5700},
  pmid      = {26888586},
  pmcid     = {PMC4795288},
  publisher = {American Academy of Sleep Medicine}
}

@article{espie2012sleepio,
  author    = {Espie, Colin A and Kyle, Simon D and Williams, Chris and Ong, Jason C and Douglas, Neil J and Hames, Peter and Brown, June S. L.},
  title     = {A Randomized, Placebo-Controlled Trial of Online Cognitive Behavioral Therapy for Chronic Insomnia Disorder Delivered via an Automated Media-Rich Web Application},
  journal   = {Sleep},
  year      = {2012},
  month     = jun,
  day       = {1},
  volume    = {35},
  number    = {6},
  pages     = {769--781},
  doi       = {10.5665/sleep.1872},
  pmid      = {22654196},
  pmcid     = {PMC3353040}
}

@article{sadhu2023dashboard,
  author    = {Sadhu, Shehjar and Solanki, Dhaval and Brick, Leslie and Nugent, Nicole and Mankodiya, Kunal},
  title     = {Designing a Clinician-Centered Wearable Data Dashboard (CarePortal): Participatory Design Study},
  journal   = {JMIR Formative Research},
  year      = {2023},
  month     = dec,
  day       = {5},
  volume    = {7},
  eid       = {e46866},
  doi       = {10.2196/46866},
  url       = {https://doi.org/10.2196/46866},
  publisher = {JMIR Publications}
}

@article{tiase2020integration,
  author    = {Tiase, Vincent L and Hull, William and McFarland, Mary M and Sward, Katherine A and Del Fiol, Guilherme and Staes, Catherine and Weir, Charlene and Cummins, Melissa R},
  title     = {Patient-Generated Health Data and Electronic Health Record Integration: A Scoping Review},
  journal   = {JAMIA Open},
  year      = {2020},
  month     = dec,
  day       = {5},
  volume    = {3},
  number    = {4},
  pages     = {619--627},
  doi       = {10.1093/jamiaopen/ooaa052},
  pmid      = {33758798},
  pmcid     = {PMC7969964}
}

@article{clegg2023real,
  title={Real, misreported, and backfilled adherence with paper sleep diaries},
  author={Clegg-Kraynok, Megan and Barnovsky, Lauren and Zhou, Eric S},
  journal={Sleep Medicine},
  volume={107},
  pages={31--35},
  year={2023},
  publisher={Elsevier}
}

@article{kristbergsdottir2023evaluating,
  title={Evaluating user compliance in mobile health apps: Insights from a 90-day study using a digital sleep diary},
  author={Kristbergsdottir, Hl{\'\i}n and Schmitz, Lisa and Arnardottir, Erna Sif and Islind, Anna Sigridur},
  journal={Diagnostics},
  volume={13},
  number={18},
  pages={2883},
  year={2023},
  publisher={MDPI}
}

@article{ibanez2018survey,
  title={A survey on sleep questionnaires and diaries},
  author={Ib{\'a}{\~n}ez, Vanessa and Silva, Josep and Cauli, Omar},
  journal={Sleep medicine},
  volume={42},
  pages={90--96},
  year={2018},
  publisher={Elsevier}
}

@article{thurman2018individual,
  author    = {Thurman, Steven M and Wasylyshyn, Nick and Roy, Heather and Lieberman, Gregory and Garcia, Javier O and Asturias, Alex and Okafor, Gold N and Elliott, James C and Giesbrecht, Barry and Grafton, Scott T and Mednick, Sara C and Vettel, Jean M},
  title     = {Individual Differences in Compliance and Agreement for Sleep Logs and Wrist Actigraphy: A Longitudinal Study of Naturalistic Sleep in Healthy Adults},
  journal   = {PLoS One},
  year      = {2018},
  month     = jan,
  day       = {29},
  volume    = {13},
  number    = {1},
  pages     = {e0191883},
  doi       = {10.1371/journal.pone.0191883},
  pmid      = {29377925},
  pmcid     = {PMC5788380},
  publisher = {Public Library of Science}
}

@article{arnardottir2022sleep,
  author    = {Arnardottir, Erna S and Islind, Anna Sigridur and \'Oskarsdottir, Mar\'ia and \'Olafsdottir, Krist\'in A and August, Elias and Jonasdottir, L\'ara and Hrubos-Str{\o}m, Harald and Saavedra, Jose M and Grote, Ludger and Hedner, Jan and H{\"o}skuldsson, Snorri and {\'A}g{\'u}stsson, Jon S and J{\'o}hannsd{\'o}ttir, Kristbj{\"o}rg R and McNicholas, Walter T and Pevernagie, Dirk and Sund, Reijo and T{\"o}yr{\"a}s, Juha and Lepp{\"a}nen, Tuomo and Sleep Revolution},
  title     = {The Sleep Revolution Project: The Concept and Objectives},
  journal   = {Journal of Sleep Research},
  year      = {2022},
  month     = aug,
  volume    = {31},
  number    = {4},
  pages     = {e13630},
  doi       = {10.1111/jsr.13630},
  pmid      = {35770626},
  publisher = {Wiley Online Library},
}

@article{silva2007relationship,
  title={Relationship between reported and measured sleep times: the sleep heart health study (SHHS)},
  author={Silva, Graciela E and Goodwin, James L and Sherrill, Duane L and Arnold, Jean L and Bootzin, Richard R and Smith, Terry and Walsleben, Joyce A and Baldwin, Carol M and Quan, Stuart F},
  journal={Journal of Clinical Sleep Medicine},
  volume={3},
  number={6},
  pages={622--630},
  year={2007},
  publisher={American Academy of Sleep Medicine}
}

@inproceedings{schmitz2022towards,
  title={Towards a Digital Sleep Diary Standard.},
  author={Schmitz, Lisa and Sveinbjarnarson, Bjarki Freyr and Gunnarsson, Gu{\dh}ni Nathan and Dav{\'\i}{\dh}sson, {\'O}lafur Andri and Dav{\'\i}{\dh}sson, {\TH}{\'o}r Breki and Arnard{\'o}ttir, Erna Sif and {\'O}skarsd{\'o}ttir, Mar{\'\i}a and Islind, Anna Sigr{\'\i}{\dh}ur},
  booktitle={AMCIS},
  volume={100},
  pages={S305},
  year={2022}
}

@article{sateia2014international,
  title={International classification of sleep disorders},
  author={Sateia, Michael J},
  journal={Chest},
  volume={146},
  number={5},
  pages={1387--1394},
  year={2014},
  publisher={Elsevier}
}

@article{ancoli2003role,
  title={The role of actigraphy in the study of sleep and circadian rhythms},
  author={Ancoli-Israel, Sonia and Cole, Roger and Alessi, Cathy and Chambers, Mark and Moorcroft, William and Pollak, Charles P},
  journal={Sleep},
  volume={26},
  number={3},
  pages={342--392},
  year={2003},
  doi={10.1093/sleep/26.3.342}
}

@article{buysse2006recommendations,
  title={Recommendations for a standard research assessment of insomnia},
  author={Buysse, Daniel J and Ancoli-Israel, Sonia and Edinger, Jack D and Lichstein, Kenneth L and Morin, Charles M},
  journal={Sleep},
  volume={29},
  number={9},
  pages={1155--1173},
  year={2006},
  publisher={Oxford University Press}
}

@article{kushida2001comparison,
  title={Comparison of actigraphic, polysomnographic, and subjective assessment of sleep parameters in sleep-disordered patients},
  author={Kushida, Clete A and Chang, Angela and Gadkary, Chitra and Guilleminault, Christian and Carrillo, Oscar and Dement, William C},
  journal={Sleep Medicine},
  volume={2},
  number={5},
  pages={389--396},
  year={2001},
  doi={10.1016/S1389-9457(00)00098-8}
}

@article{edinger2021behavioral,
  title={Behavioral and psychological treatments for chronic insomnia disorder in adults: An American Academy of Sleep Medicine clinical practice guideline},
  author={Edinger, Jack D and Arnedt, J Todd and Bertisch, Sarah M and Carney, Colleen E and Harrington, Kelly E and Lichstein, Kenneth L and Sateia, Michael J and Troxel, Wendy M and and others},
  journal={Journal of Clinical Sleep Medicine},
  year={2021},
  volume={17},
  number={2},
  pages={255--262},
  doi={10.5664/jcsm.8986}
}

@article{islind2019shift,
  title={Shift in translations: Data work with patient-generated health data in clinical practice},
  author={Islind, Anna Sigridur and Lindroth, Tomas and Lundin, Johan and Steineck, Gunnar},
  journal={Health informatics journal},
  volume={25},
  number={3},
  pages={577--586},
  year={2019},
  publisher={SAGE Publications Sage UK: London, England}
}

@article{cerna2020changing,
  title={Changing categorical work in healthcare: the use of patient-generated health data in cancer rehabilitation},
  author={Cerna, Katerina and Grisot, Miria and Islind, Anna Sigridur and Lindroth, Tomas and Lundin, Johan and Steineck, Gunnar},
  journal={Computer Supported Cooperative Work (CSCW)},
  volume={29},
  number={5},
  pages={563--586},
  year={2020},
  publisher={Springer}
}

@article{morin2009cognitive,
  title={Cognitive Behavioral Therapy, Singly and Combined With Medication, for Persistent Insomnia: A Randomized Controlled Trial},
  author={Morin, Charles M and Valli{\`e}res, Annie and Guay, Bernard and Ivers, Hans and Savard, Jos{\'e}e and M{\'e}rette, Chantal and Bastien, C{\'e}lyne and Baillargeon, Lucie},
  journal={JAMA},
  volume={301},
  number={19},
  pages={2005--2015},
  year={2009},
  doi={10.1001/jama.2009.682}
}

@book{perlis2005cognitive,
  title={Cognitive Behavioral Treatment of Insomnia: A Session by Session Guide},
  author={Perlis, Michael L and Jungquist, Carla and Smith, Michael T and Posner, David},
  year={2005},
  publisher={Springer Science and Business Media}
}

@article{furukawa2024components,
  title={Components and Delivery Formats of Cognitive Behavioral Therapy for Chronic Insomnia in Adults: A Systematic Review and Component Network Meta Analysis},
  author={Furukawa, Toshi A and Zhou, Xinyu and Hino, Kohei and Watanabe, Naoki and Shinohara, Katsuaki and Murayama, Hiroshi and Jones, Candace M and Salanti, Georgia and Cuijpers, Pim},
  journal={JAMA Psychiatry},
  year={2024},
  volume={81},
  number={2},
  pages={166--176},
  doi={10.1001/jamapsychiatry.2023.4227}
}

@article{benjafield2025estimation,
  author    = {Benjafield, Adam V and Kuniyoshi, Fatima H. Sert and Malhotra, Atul and Martin, Jennifer L and Morin, Charles M and Maurer, Leonie F and Cistulli, Peter A and P{\'e}pin, Jean-Louis and Wickwire, Emerson M and medXcloud group},
  title     = {Estimation of the Global Prevalence and Burden of Insomnia: A Systematic Literature Review-Based Analysis},
  journal   = {Sleep Medicine Reviews},
  year      = {2025},
  month     = aug,
  pages     = {102121},
  volume    = {82},
  doi       = {10.1016/j.smrv.2025.102121},
  pmid      = {40627924},
  publisher = {Elsevier},
}

@article{rossman2019cognitive,
  title={Cognitive-behavioral therapy for insomnia: an effective and underutilized treatment for insomnia},
  author={Rossman, Jeffrey},
  journal={American journal of lifestyle medicine},
  volume={13},
  number={6},
  pages={544--547},
  year={2019},
  publisher={SAGE Publications Sage CA: Los Angeles, CA}
}

@article{van2019cognitive,
  title={Cognitive behavioral therapy for insomnia: a meta-analysis of long-term effects in controlled studies},
  author={van der Zweerde, Tanja and Bisdounis, Lampros and Kyle, Simon D and Lancee, Jaap and van Straten, Annemieke},
  journal={Sleep medicine reviews},
  volume={48},
  pages={101208},
  year={2019},
  publisher={Elsevier}
}

@article{anwar2025effect,
  title={Effect of Cognitive Behavioral Therapy for Insomnia in Patients With Co-Morbid Insomnia and Sleep Apnea: A Systematic Review and Meta-Analysis of Randomized Controlled Trials},
  author={Anwar, Deva Fitra Firdausa and Fidiana, Fidiana and Irwadi, Irfiansyah and Rosyid, Alfian Nur and Salma, Zaskia Nafisa and Hibatulloh, Muhammad Farhan},
  journal={Sleep Medicine Research},
  volume={16},
  number={1},
  pages={18--32},
  year={2025},
  publisher={Korean Society of Sleep Medicine}
}

@misc{resmed2025Global,
	author = {Resmed},
	title = {2025 {G}lobal {S}leep {S}urvey | A World Struggling With Poor Sleep},
	howpublished = {\url{https://sleepsurvey.resmed.com/}},
	year = {},
	note = {[Accessed 05-09-2025]},
}

@inproceedings{xu2025exploring,
  title={Exploring the Use of Robots for Diary Studies},
  author={Xu, Michael F and Mutlu, Bilge},
  booktitle={2025 20th ACM/IEEE International Conference on Human-Robot Interaction (HRI)},
  pages={174--182},
  year={2025},
  organization={IEEE}
}

@article{johnson2018environmental,
  title={Environmental determinants of insufficient sleep and sleep disorders: implications for population health},
  author={Johnson, Dayna A and Billings, Martha E and Hale, Lauren},
  journal={Current epidemiology reports},
  volume={5},
  number={2},
  pages={61--69},
  year={2018},
  publisher={Springer}
}

@article{billings2020physical,
  title={Physical and social environment relationship with sleep health and disorders},
  author={Billings, Martha E and Hale, Lauren and Johnson, Dayna A},
  journal={Chest},
  volume={157},
  number={5},
  pages={1304--1312},
  year={2020},
  publisher={Elsevier}
}

@article{chang2020impact,
  title={The impact of the environment on the quality of life and the mediating effects of sleep and stress},
  author={Chang, Katherine Ka Pik and Wong, Frances Kam Yuet and Chan, Ka Long and Wong, Fiona and Ho, Hung Chak and Wong, Man Sing and Ho, Yuen Shan and Yuen, John Wai Man and Siu, Judy Yuen-man and Yang, Lin},
  journal={International Journal of Environmental Research and Public Health},
  volume={17},
  number={22},
  pages={8529},
  year={2020},
  publisher={MDPI}
}

@article{irish2015role,
  title={The role of sleep hygiene in promoting public health: A review of empirical evidence},
  author={Irish, Leah A and Kline, Christopher E and Gunn, Heather E and Buysse, Daniel J and Hall, Martica H},
  journal={Sleep medicine reviews},
  volume={22},
  pages={23--36},
  year={2015}
}

@article{wei2024leveraging,
  title={Leveraging large language models to power chatbots for collecting user self-reported data},
  author={Wei, Jing and Kim, Sungdong and Jung, Hyunhoon and Kim, Young-Ho},
  journal={Proceedings of the ACM on Human-Computer Interaction},
  volume={8},
  number={CSCW1},
  pages={1--35},
  year={2024},
  publisher={ACM New York, NY, USA}
}

@inproceedings{mahmood2025voice,
  title={Voice assistants for health self-management: Designing for and with older adults},
  author={Mahmood, Amama and Cao, Shiye and Stiber, Maia and Antony, Victor Nikhil and Huang, Chien-Ming},
  booktitle={Proceedings of the 2025 CHI Conference on Human Factors in Computing Systems},
  pages={1--22},
  year={2025}
}

@article{yang2024talk2care,
  title={Talk2care: An llm-based voice assistant for communication between healthcare providers and older adults},
  author={Yang, Ziqi and Xu, Xuhai and Yao, Bingsheng and Rogers, Ethan and Zhang, Shao and Intille, Stephen and Shara, Nawar and Gao, Guodong Gordon and Wang, Dakuo},
  journal={Proceedings of the ACM on Interactive, Mobile, Wearable and Ubiquitous Technologies},
  volume={8},
  number={2},
  pages={1--35},
  year={2024},
  publisher={ACM New York, NY, USA}
}

@inproceedings{kim2024mindfuldiary,
  author    = {Kim, Taewan and Bae, Seolyeong and Kim, Hyun Ah and Lee, Su-Woo and Hong, Hwajung and Yang, Chanmo and Kim, Young-Ho},
  title     = {MindfulDiary: Harnessing Large Language Model to Support Psychiatric Patients' Journaling},
  booktitle = {Proceedings of the 2024 CHI Conference on Human Factors in Computing Systems},
  year      = {2024},
  publisher = {Association for Computing Machinery},
  address   = {New York, NY, USA},
  location  = {Honolulu, HI, USA},
  series    = {CHI '24},
  articleno = {701},
  numpages  = {20},
  isbn      = {9798400703300},
  doi       = {10.1145/3613904.3642937}
}

@article{fang2024physiollm,
  author    = {Fang, Cathy Mengying and Danry, Valdemar and Whitmore, Nathan and Bao, Andria and Hutchison, Andrew and Pierce, Cayden and Maes, Pattie},
  title     = {PhysioLLM: Supporting Personalized Health Insights with Wearables and Large Language Models},
  journal   = {arXiv preprint arXiv:2406.19283},
  year      = {2024},
  month     = jun,
  doi       = {10.48550/arXiv.2406.19283},
  url       = {https://arxiv.org/abs/2406.19283}
}

@inproceedings{brooke1996sus,
  author    = {Brooke, John},
  title     = {SUS: A Quick and Dirty Usability Scale},
  booktitle = {Usability Evaluation in Industry},
  year      = {1996},
  editor    = {Jordan, Patrick W and Thomas, Bruce and Weerdmeester, Ian and McClelland, Bernard},
  pages     = {189--194},
  publisher = {Taylor \& Francis},
  address   = {London}
}

@article{bangor2009determining,
  title={Determining what individual SUS scores mean: Adding an adjective rating scale},
  author={Bangor, Aaron and Kortum, Philip and Miller, James},
  journal={Journal of usability studies},
  volume={4},
  number={3},
  pages={114--123},
  year={2009},
  publisher={Usability Professionals' Association Bloomingdale, IL}
}

@article{koubaa2023exploring,
  author    = {Koubaa, Anis and Boulila, Wadii and Ghouti, Lahouari and Alzahem, Ayyub and Latif, Shahid},
  title     = {Exploring ChatGPT Capabilities and Limitations: A Critical Review of the NLP Game Changer},
  journal   = {Preprints},
  year      = {2023},
  month     = mar,
  day       = {27},
  doi       = {10.20944/preprints202303.0438.v1},
}

@article{ong2024ethical,
  author    = {Ong, Jasmine Chiat Ling and Chang, Shelley Yin-Hsi and William, Wasswa and Butte, Atul J and Shah, Nigam H and Chew, Lita Sui Tjien and Liu, Nan and Doshi-Velez, Finale and Lu, Wei and Savulescu, Julian and Ting, Daniel Shu Wei},
  title     = {Ethical and Regulatory Challenges of Large Language Models in Medicine},
  journal   = {The Lancet Digital Health},
  year      = {2024},
  month     = jun,
  volume    = {6},
  number    = {6},
  pages     = {e428--e432},
  doi       = {10.1016/S2589-7500(24)00061-X},
  pmid      = {38658283},
  publisher = {Elsevier}
}

@article{hastings2024preventing,
  title={Preventing harm from non-conscious bias in medical generative AI},
  author={Hastings, Janna},
  journal={The Lancet Digital Health},
  volume={6},
  number={1},
  pages={e2--e3},
  year={2024},
  publisher={Elsevier}
}

@article{mustafa2005sleep,
  title={Sleep problems and the risk for sleep disorders in an outpatient veteran population},
  author={Mustafa, Masroor and Erokwu, Nkolika and Ebose, Idowa and Strohl, Kingman},
  journal={Sleep and Breathing},
  volume={9},
  number={2},
  pages={57--63},
  year={2005},
  publisher={Springer}
}

@article{ong2018concept,
  title={A Concept Map of Behavioral Sleep Medicine: Defining the Scope of the Field and Strategic Priorities},
  author={Ong, Jason C and Arand, Donna and Schmitz, Mark and Baron, Kelly and Blackburn, Robert and Grandner, Michael A and Lichstein, Kenneth L and Nowakowski, Sara and Teixeira, Carla and Boling, Kelly and Dawson, Stacy C and Hansen, Kim},
  journal={Behavioral Sleep Medicine},
  volume={16},
  number={6},
  pages={523--526},
  year={2018},
  publisher={Taylor \& Francis},
  doi={10.1080/15402002.2018.1507672}
}

@article{sinha2022adherence,
  title={Adherence and Engagement With a Cognitive Behavioral Therapy--Based Conversational Agent (Wysa for Chronic Pain) Among Adults With Chronic Pain: Survival Analysis},
  author={Sinha, Chaitali and Cheng, Abby L and Kadaba, Madhura},
  journal={JMIR Formative Research},
  year={2022},
  doi={10.2196/37302}
}

@article{drake2016promise,
  title={The promise of digital CBT-I},
  author={Drake, Christopher L},
  journal={Sleep},
  volume={39},
  number={1},
  pages={13--14},
  year={2016},
  publisher={Oxford University Press}
}

@article{sanabria2025active,
  author    = {Sanabria, Alicia Salamanca and Fogel, Anna and Natarajan, Padmapriya and Rodriguez, Alina and Gunnar, Eriksson Johan},
  title     = {Active Components in Digital Health Interventions for Sleep among Adolescents: A Systematic Review and Meta-Analysis of Randomized Controlled Trials},
  journal   = {Research Square},
  year      = {2025},
  month     = may,
  day       = {7},
  doi       = {10.21203/rs.3.rs-6609677/v1},
  url       = {https://doi.org/10.21203/rs.3.rs-6609677/v1},
  note      = {Preprint}
}

@incollection{bootzin1972behavioral,
  title        = {Stimulus control treatment for insomnia},
  author       = {Bootzin, Richard R.},
  booktitle    = {Behavior Modification: An Introductory Textbook},
  editor       = {Hersen, Michel and Eisler, Richard M. and Miller, Paul M.},
  publisher    = {Academic Press},
  address      = {New York},
  year         = {1972},
  pages        = {pp. 295--308}
}

@article{maharjan2022difference,
  author    = {Maharjan, Raju and Rohani, Darius A. and Doherty, Kevin and B{\ae}kgaard, Per and Bardram, Jakob E.},
  title     = {What Is the Difference? Investigating the Self-Report of Wellbeing via Conversational Agent and Web App},
  journal   = {IEEE Pervasive Computing},
  year      = {2022},
  volume    = {21},
  number    = {2},
  pages     = {60--68},
  doi       = {10.1109/MPRV.2022.3147374}
}

@inproceedings{li2024diaryhelper,
  author       = {Li, Junze and He, Changyang and Hu, Jiaxiong and Jia, Boyang and Halevy, Alon Y. and Ma, Xiaojuan},
  title        = {DiaryHelper: Exploring the Use of an Automatic Contextual Information Recording Agent for Elicitation Diary Study},
  booktitle    = {Proceedings of the 2024 CHI Conference on Human Factors in Computing Systems},
  series       = {CHI '24},
  year         = {2024},
  location     = {Honolulu, HI, USA},
  publisher    = {Association for Computing Machinery},
  address      = {New York, NY, USA},
  articleno    = {818},
  numpages     = {16},
  isbn         = {979-8-4007-0330-0},
  doi          = {10.1145/3613904.3642853},
  url          = {https://doi.org/10.1145/3613904.3642853}
}

@article{bandyopadhyay2024strengths,
  author    = {Bandyopadhyay, Anuja and Oks, Margarita and Sun, Haoqi and Prasad, Bharati and Rusk, Sam and Jefferson, Felicia and Malkani, Roneil Gopal and Haghayegh, Shahab and Sachdeva, Ramesh and Hwang, Dennis and Agustsson, Jon and Mignot, Emmanuel and Summers, Michael and Fabbri, Daniele and Deak, Mihaly and Anastasi, Maria and Sampson, Andrew and Van Hout, Steven and Seixas, Azizi},
  title     = {Strengths, Weaknesses, Opportunities, and Threats of Using AI-Enabled Technology in Sleep Medicine: A Commentary},
  journal   = {Journal of Clinical Sleep Medicine},
  year      = {2024},
  month     = jul,
  day       = {1},
  volume    = {20},
  number    = {7},
  pages     = {1183--1191},
  doi       = {10.5664/jcsm.11132},
  pmid      = {38533757},
  pmcid     = {PMC11217619},
  publisher = {American Academy of Sleep Medicine}
}

@article{qaseem2016management,
  title={Management of chronic insomnia disorder in adults: a clinical practice guideline from the American College of Physicians},
  author={Qaseem, Amir and Kansagara, Devan and Forciea, Mary Ann and Cooke, Molly and Denberg, Thomas D and Clinical Guidelines Committee of the American College of Physicians*},
  journal={Annals of internal medicine},
  volume={165},
  number={2},
  pages={125--133},
  year={2016},
  publisher={American College of Physicians}
}

@article{taylor2017internet,
  title={Internet and in-person cognitive behavioral therapy for insomnia in military personnel: a randomized clinical trial},
  author={Taylor, Daniel J and Peterson, Alan L and Pruiksma, Kristi E and Young-McCaughan, Stacey and Nicholson, Karin and Mintz, Jim and STRONG STAR Consortium Borah Elisa V. PhD Dondanville Katherine A. PsyD Hale Willie J. PhD Litz Brett T. PhD Roache John D. PhD Borah Dr},
  journal={Journal of Sleep and Sleep Disorders Research},
  volume={40},
  number={6},
  pages={zsx075},
  year={2017},
  publisher={Oxford University Press US}
}

@misc{sivertsen2013future,
  title={The future of insomnia treatment—the challenge of implementation},
  author={Sivertsen, B{\o}rge and Vedaa, {\O}ystein and Nordgreen, Tine},
  journal={Sleep},
  volume={36},
  number={3},
  pages={303--304},
  year={2013},
  publisher={Oxford University Press}
}

@article{thomas2016behavioral,
  title={Where are the behavioral sleep medicine providers and where are they needed? A geographic assessment},
  author={Thomas, Arthur and Grandner, Michael and Nowakowski, Sara and Nesom, Genevieve and Corbitt, Charles and Perlis, Michael L},
  journal={Behavioral sleep medicine},
  volume={14},
  number={6},
  pages={687--698},
  year={2016},
  publisher={Taylor \& Francis}
}

@misc{perlis2008can,
  title={How can we make CBT-I and other BSM services widely available?},
  author={Perlis, Michael L and Smith, Michael T},
  journal={Journal of Clinical Sleep Medicine},
  volume={4},
  number={1},
  pages={11--13},
  year={2008},
  publisher={American Academy of Sleep Medicine}
}

@article{germain2021enhancing,
  author    = {Germain, Anne and Markwald, Rachel R and King, Erika and Bramoweth, Adam D and Wolfson, Megan and Seda, Gilbert and Han, Tony and Miggantz, Erin and O’Reilly, Brian and Hungerford, Lars and Sitzer, Traci and Mysliwiec, Vincent and Hout, Joseph and Wallace, Meredith},
  title     = {Enhancing Behavioral Sleep Care with Digital Technology: Study Protocol for a Hybrid Type 3 Implementation-Effectiveness Randomized Trial},
  journal   = {Trials},
  year      = {2021},
  month     = jan,
  day       = {11},
  volume    = {22},
  number    = {1},
  pages     = {46},
  doi       = {10.1186/s13063-020-04974-z},
  publisher = {Springer}
}

@article{arigo2019history,
  title={The history and future of digital health in the field of behavioral medicine},
  author={Arigo, Danielle and Jake-Schoffman, Danielle E and Wolin, Kathleen and Beckjord, Ellen and Hekler, Eric B and Pagoto, Sherry L},
  journal={Journal of behavioral medicine},
  volume={42},
  number={1},
  pages={67--83},
  year={2019},
  publisher={Springer}
}

@article{aan2012mood,
  title={Mood disorders in everyday life: A systematic review of experience sampling and ecological momentary assessment studies},
  author={aan het Rot, Marije and Hogenelst, Koen and Schoevers, Robert A},
  journal={Clinical psychology review},
  volume={32},
  number={6},
  pages={510--523},
  year={2012},
  publisher={Elsevier}
}

@article{stone2003patient,
  title={Patient compliance with paper and electronic diaries},
  author={Stone, Arthur A and Shiffman, Saul and Schwartz, Joseph E and Broderick, Joan E and Hufford, Michael R},
  journal={Controlled clinical trials},
  volume={24},
  number={2},
  pages={182--199},
  year={2003},
  publisher={Elsevier}
}

@article{burke2011self,
  title={Self-monitoring in weight loss: a systematic review of the literature},
  author={Burke, Lora E and Wang, Jing and Sevick, Mary Ann},
  journal={Journal of the American Dietetic Association},
  volume={111},
  number={1},
  pages={92--102},
  year={2011},
  publisher={Elsevier}
}

@article{shiffman2009ecological,
  title={Ecological momentary assessment (EMA) in studies of substance use.},
  author={Shiffman, Saul},
  journal={Psychological assessment},
  volume={21},
  number={4},
  pages={486},
  year={2009},
  publisher={American Psychological Association}
}

@article{nes2013web,
  title={Web-based, self-management enhancing interventions with e-diaries and personalized feedback for persons with chronic illness: A tale of three studies},
  author={Nes, Andrea AG and Eide, Hilde and Kristj{\'a}nsd{\'o}ttir, {\'O}l{\"o}f Birna and van Dulmen, Sandra},
  journal={Patient education and counseling},
  volume={93},
  number={3},
  pages={451--458},
  year={2013},
  publisher={Elsevier}
}

@article{aarsand2010mobile,
  title={Mobile phone-based self-management tools for type 2 diabetes: the few touch application},
  author={{\AA}rsand, Eirik and Tatara, Naoe and {\O}stengen, Geir and Hartvigsen, Gunnar},
  journal={Journal of diabetes science and technology},
  volume={4},
  number={2},
  pages={328--336},
  year={2010},
  publisher={SAGE Publications}
}

@article{hunt2015technology,
  title={Technology and diabetes self-management: an integrative review},
  author={Hunt, Caralise W},
  journal={World journal of diabetes},
  volume={6},
  number={2},
  pages={225},
  year={2015}
}

@article{van1999patient,
  title={A patient diary as a tool to improve medicine compliance},
  author={van Berge Henegouwen, MTH and Van Driel, HF and Kasteleijn-nolst trenit{\'e}, DGA},
  journal={Pharmacy World and Science},
  volume={21},
  number={1},
  pages={21--24},
  year={1999},
  publisher={Springer}
}

@article{tang2013online,
  author    = {Tang, Paul C and Overhage, J. Marc and Chan, Albert Solomon and Brown, Nancy L and Aghighi, Bahar and Entwistle, Martin P and Hui, Siu Lui and Hyde, Shauna M and Klieman, Linda H and Mitchell, Charlotte J and Perkins, Ann J and Qureshi, Laila S and Waltimyer, Timothy A and Winters, Lawrence J and Young, Christopher Y},
  title     = {Online Disease Management of Diabetes: Engaging and Motivating Patients Online with Enhanced Resources-Diabetes (EMPOWER-D), a Randomized Controlled Trial},
  journal   = {Journal of the American Medical Informatics Association},
  year      = {2013},
  month     = may,
  day       = {1},
  volume    = {20},
  number    = {3},
  pages     = {526--534},
  doi       = {10.1136/amiajnl-2012-001263},
  pmid      = {23171659},
  pmcid     = {PMC3628059},
  publisher = {BMJ Publishing Group}
}

@article{abbasian2025conversational,
  title={Conversational health agents: a personalized large language model-powered agent framework},
  author={Abbasian, Mahyar and Azimi, Iman and Rahmani, Amir M and Jain, Ramesh},
  journal={JAMIA Open},
  volume={8},
  number={4},
  pages={ooaf067},
  year={2025},
  publisher={Oxford University Press}
}

@article{merrill2024transforming,
  author    = {Merrill, Mike A and Paruchuri, Akshay and Rezaei, Naghmeh and Kovacs, Geza and Perez, Javier and Liu, Yun and Schenck, Erik and Hammerquist, Nova and Sunshine, Jake and Tailor, Shyam and Lad, Vijay and Jopling, Jacob and Yuan, Minghao and Bozkurt, Arda and Bansal, Mohit and Choudhury, Munmun and Denecke, Kerstin and Sharma, Amit and Lee, Michael and Shah, Nigam H and Althoff, Tim},
  title     = {Transforming Wearable Data into Health Insights Using Large Language Model Agents},
  journal   = {arXiv preprint arXiv:2406.06464},
  year      = {2024},
  month     = jun,
  doi       = {10.48550/arXiv.2406.06464},
  url       = {https://arxiv.org/abs/2406.06464},
}

@book{altevogt2006sleep,
  author    = {Colten, Harvey R and Altevogt, Bruce M},
  title     = {Sleep Disorders and Sleep Deprivation: An Unmet Public Health Problem},
  year      = {2006},
  publisher = {National Academies Press},
  address   = {Washington, DC},
  isbn      = {9780309101110},
  pmid      = {20669438},
  editor    = {{Institute of Medicine (US) Committee on Sleep Medicine and Research}}
}

@article{shah2025effects,
  author    = {Shah, Arambh Sanjay and Pant, Mitresh Raj and Bommasamudram, Tulasiram and Nayak, Kirtana Raghurama and Roberts, Spencer S. H. and Gallagher, Chloe and Vaishali, K and Edwards, Ben J and Tod, David and Davis, Fiddy and Pullinger, Samuel A},
  title     = {Effects of Sleep Deprivation on Physical and Mental Health Outcomes: An Umbrella Review},
  journal   = {American Journal of Lifestyle Medicine},
  year      = {2025},
  month     = may,
  day       = {27},
  pages     = {15598276251346752},
  doi       = {10.1177/15598276251346752},
  pmid      = {40443808},
  pmcid     = {PMC12116485},
  publisher = {SAGE Publications}
}

@article{lim2023need,
  author    = {Lim, Diane C and Najafi, Arezu and Afifi, Lamia and Bassetti, Claudio L. A. and Buysse, Daniel J and Han, Fang and H{\"o}gl, Birgit and Melaku, Yohannes Adama and Morin, Charles M and Pack, Allan I and Poyares, Dalva and Somers, Virend and Eastwood, Peter and Zee, Phyllis and Jackson, Chandra},
  title     = {The Need to Promote Sleep Health in Public Health Agendas Across the Globe},
  journal   = {The Lancet Public Health},
  year      = {2023},
  month     = oct,
  volume    = {8},
  number    = {10},
  pages     = {e820--e826},
  doi       = {10.1016/S2468-2667(23)00182-2},
  publisher = {Elsevier}
}

@incollection{braun2023doing,
  title={Doing reflexive thematic analysis},
  author={Braun, Virginia and Clarke, Victoria and Hayfield, Nikki and Davey, Louise and Jenkinson, Elizabeth},
  booktitle={Supporting research in counselling and psychotherapy: Qualitative, quantitative, and mixed methods research},
  pages={19--38},
  year={2023},
  publisher={Springer}
}

@article{paranjape2023art,
  title={Art: Automatic multi-step reasoning and tool-use for large language models},
  author={Paranjape, Bhargavi and Lundberg, Scott and Singh, Sameer and Hajishirzi, Hannaneh and Zettlemoyer, Luke and Ribeiro, Marco Tulio},
  journal={arXiv preprint arXiv:2303.09014},
  year={2023}
}

@article{ahmad2023creating,
  title={Creating trustworthy llms: Dealing with hallucinations in healthcare ai},
  author={Ahmad, Muhammad Aurangzeb and Yaramis, Ilker and Roy, Taposh Dutta},
  journal={arXiv preprint arXiv:2311.01463},
  year={2023}
}

@article{akheel2025guardrails,
  title={Guardrails for large language models: A review of techniques and challenges},
  author={Akheel, S},
  journal={J Artif Intell Mach Learn \& Data Sci},
  volume={3},
  number={1},
  pages={2504--2512},
  year={2025}
}

\newpage

\appendix
\section*{Appendix}

\section{\am{Methods}}

\subsection{\am{Unstructured conversations to JSON conversion}}
\label{app:json}

\am{To validate the LLM-based diary-to-JSON conversion, we developed a coding scheme with two criteria: (1) \textbf{Completeness} (0 = absent, 1 = partially complete, 2 = fully complete), which assesses whether all required information from a diary element is represented in the JSON output, and (2) \textbf{Accuracy} (0 = inaccurate, 1 = accurate), which evaluates whether the extracted value, when present, correctly reflects what the participant reported during the conversation. Two coders independently applied this scheme to each diary element/variable listed in Table~\ref{tab:coding-questions}.}
\begin{table*}[h!]
\centering
\caption{\am{Diary questions and system-derived variables mapped to JSON output fields. All variables were validated for completeness and accuracy.}}

\label{tab:coding-questions}
\small
\begingroup

\begin{tabular}{p{10cm} p{4.5cm}}
\toprule
\textbf{Diary Element} & \textbf{Variable} \\
\midrule
\multicolumn{2}{l}{\textit{\textbf{Morning Diary Questions}}} \\
\midrule
How are you feeling today? & mood\_morning \\
What time did you get into bed last night? & time\_in\_bed \\
What time did you turn out the lights (with the intention of going to sleep)? & lights\_out\_time \\
After turning out the lights, how many minutes did it take you to fall asleep? & sleep\_latency \\
Did you go to bed earlier than planned? & on\_time \\
How many awakenings did you have last night? Please add up the total amount of time you were awake during the night. & awakenings \\
What was the time of your final awakening? & final\_awakening\_time \\
What time did you get out of bed? & out\_of\_bed\_time \\
Did you wake up earlier than planned? If yes, how much earlier? & woke\_earlier \\
Please rate your sleep quality. & sleep\_quality \\
\hspace*{1em} & sleep\_quality\_comment \\
Use of devices or sleep aids. & sleep\_aids \\
Provide any important details about your sleep last night. & additional\_comments \\

\midrule
\multicolumn{2}{l}{\textit{\textbf{Evening Diary Questions}}} \\
\midrule
How are you feeling today? & mood\_evening \\
Did you take any naps today? If yes, provide details. & nap\_details \\
Did you consume alcohol? If yes, provide type, amount, and time. & alcohol\_intake \\
Did you drink caffeinated drinks? If yes, provide type, amount, and time. & caffeinated\_intake \\
Did you take any medication? If yes, state type and time. & medication\_intake \\
Did you do any physical activity or exercise? Provide type, duration, and time. & exercise\_activity \\
Please rate your current stress level. & stress\_before\_bed \\
\hspace*{1em} & stress\_before\_bed\_comment \\
Please rate your fatigue level during the day. & fatigue\_level \\
\hspace*{1em} & fatigue\_level\_comment \\
Please describe your bedtime routine. & bedtime\_routine \\

\midrule
\multicolumn{2}{l}{\textit{\textbf{System-Derived Variables}}} \\
\midrule
Date of diary entry & date \\
Reason-focused tags derived from the diary content & tags \\

\bottomrule
\end{tabular}
\endgroup
\end{table*}

\end{document}